\documentclass[11pt]{article}

\usepackage{amsmath,amssymb}
\usepackage{graphicx,color}

\newcommand{\f}[2]{\frac{#1}{#2}}
\newcommand{\K}[1]{{\rm \bf K}\mbox{$\left({#1}\right)$}}
\newcommand{\E}[1]{{\rm \bf E}\mbox{$\left({#1}\right)$}}
\newcommand{\PP}[1]{{\bf \Pi}\mbox{$\left({#1}\right)$}}

\newcommand{\ko}[1]{\left( #1 \right)}
\newcommand{\kko}[1]{\left[ #1 \right]}
\newcommand{\kkko}[1]{\left\{ #1 \right\}}
\newcommand{\abs}[1]{\left| #1 \right|}
\newcommand{\ket}[1]{\left| #1 \right\rangle}
\newcommand{\bra}[1]{\left\langle #1 \right|}
\newcommand{\bmt}[1]{{{\mbox{\boldmath$ #1 $}}}}
\newcommand{\komoji}[1]{\mbox{$#1$}}
\newcommand{\ord}[1]{{\mathcal O\mbox{\small $\left(#1\right)$}}}

\def\A{{\mathcal A}}
\def\B{{\mathcal B}}
\def\SS{{\mathcal S}}

\def\tsig{{\widetilde\sigma}}
\def\ttau{{\widetilde\tau}}

\def\S{{S${}^5$}}

\def\AdSS{{AdS${}_5\times {}$S${}^5$}}
\def\Nf{${\mathcal N}=4$}
\def\H{{\mathcal H}}
\def\Tr{{\rm Tr\,}}

\def\L{{\mathcal L}}

\def\cJ{{\mathcal{J}}}

\def\R{{\mathbb{R}}}

\def\N{{\mathcal N}}
\def\C{{\mathcal C}}
\def\A{{\mathcal A}}
\def\pa{\partial}
\def\eq{\equiv}
\def\be{\beta}
\def\al{\alpha}
\def\ga{\gamma}
\def\vp{\varphi}

\def\hg{\hat\gamma}
\def\hs{\hat\sigma}

\def\no{\nonumber}
\def\xp{{x}_{+}}
\def\xm{{x}_{-}}
\def\xz{{x}_{0}}
\def\k{{k}}

\numberwithin{equation}{section}
\begin{document}
\quad 
\vspace{-2.0cm}

\begin{flushright}
\parbox{4cm}
{DAMTP-2006-04 \hfill \\
UT-05-20 \hfill \\
{\tt hep-th/0601109}\hfill
 }
\end{flushright}

\vspace*{0.5cm}

\begin{center}
\Large\bf 
The Anatomy of Gauge/String Duality in Lunin-Maldacena Background
\end{center}
\vspace{1.0cm}

\centerline{\large Heng-Yu Chen${}^{\dagger}$ and Keisuke Okamura${}^{\ddagger}$}
\vspace{0.6cm}

\begin{center}
${}^{\dagger}$\emph{DAMTP, Centre for Mathematical Sciences, Cambridge University,\\
Wilberforce Road, Cambridge CB3 OWA, UK.} \\
\vspace{0.5cm}
${}^{\ddagger}$\emph{Department of Physics, Faculty of Science, University of Tokyo,\\
Bunkyo-ku, Tokyo 113-0033, Japan.} \\
\vspace*{0.3cm}
${}^{\dagger}${\tt h.y.chen@damtp.cam.ac.uk}\qquad 
${}^{\ddagger}${\tt okamura@hep-th.phys.s.u-tokyo.ac.jp}

\end{center}

\vspace*{0.7cm}

\centerline{\bf Abstract} 

\vspace*{0.5cm}

We consider the correspondence between the spinning string solutions in Lunin-Maldacena background and the single trace operators in the Leigh-Strassler deformation of {\Nf} SYM.
By imposing an appropriate rotating string ans\"atz on the Landau-Lifshitz reduced sigma model in the deformed $SU(2)$ sector, we find two types of `elliptic' solutions with two spins, which turn out to be the solutions associated with the Neumann-Rosochatius system.
We then calculate the string energies as functions of spins, and obtain their explicit forms in terms of a set of moduli parameters.
On the deformed spin-chain side, we explicitly compute the one-loop anomalous dimensions of the gauge theory operators dual to each of the two types of spinning string solutions, extending and complementing the results of hep-th/0511164.
Moreover, we propose explicit ans\"atz on how the locations of the Bethe strings are affected due to the deformation, with several supports from the string side.

\vspace*{1.0cm}

\vfill

\thispagestyle{empty}
\setcounter{page}{0}

\newpage 

\section{Introduction}

The conjectured duality between string and gauge theories represents
one of the major breakthroughs in 
theoretical high energy physics research for the past decade 
\cite{Maldacena:1997re,Witten:1998qj,Gubser:1998bc}. 
However, as common for most dualities, the exact analytic information
for the two dual theories 
are usually only available in the complementary regions of their
coupling space, 
namely the duality is of a `weak/strong' type.
For example in AdS/CFT, on the gauge side, the 't Hooft coupling $\lambda$ is sufficiently small, whereas on the string side, $\lambda$ has to be large for ignoring the quantum corrections (in $\al'$ sense). This feature renders it a very difficult task to explicitly prove the equivalence of the two theories for their entire coupling space.

On the other hand, when certain other parameters (``quantum numbers'') beside $\lambda$ become large, 
the structures of the two theories usually simplify, and 
explicit tests for the duality can become available. 
The pioneering work by Berenstein, Maldacena and Nastase (BMN)
\cite{Berenstein:2002jq} is a classic example 
where the angular momentum (or ``spin'') $J$ of the string state becomes large and 
it allows us to test the AdS/CFT as a realization of the gauge/string
duality in the full stringy sense. Moreover, 
such large quantum numbers usually allow us to perform
certain semi-classical calculations, 
where the quantum corrections are usually suppressed \cite{Gubser:2002tv,Frolov:2002av}. 

The recent emergence of integrable structures in both
${\mathcal{N}}=4$ SYM and IIB string theory 
in {\AdSS} took these ideas further. 
Began with the remarkable observation made in \cite{Minahan:2002ve} by Minahan and Zarembo, 
the problem of computing the one-loop anomalous dimensions of ``long''
single trace gauge invariant operators 
in ${\mathcal{N}}=4$ SYM
can be rewritten to the diagonalization of certain integrable spin chain
Hamiltonians. 
More specifically, this involves solving a set of Bethe ans\"atz
equations, and in so-called ``scaling'' (or ``thermodynamic'') limit, 
where the number of sites of the spin chain becomes large, the problem further translates into the well-known Riemann-Hilbert problem \cite{Beisert:2003xu}.
In parallel, it was realized that for the semi-classical strings carrying 
one or more large angular momenta and propogating in {\AdSS}, the $\al'$-corrections to their energies are suppressed at one-loop in derivative expansion \cite{Frolov:2003qc}
(See \cite{Tseytlin:2003ii,Tseytlin:2004xa} for comprehensive
reviews). 

Moreover, the energy expression for the semi-classical string, when expanded in effective coupling $\lambda/J^{2}$, has the required analyticity for it to be identified with the perturbative expansion for the anomalous dimension of the dual gauge theory operator. 
The expansion coefficients from the independent gauge and string theoretic calculations showed striking non-trivial agreements for both near-BPS (BMN) and far-from-BPS (Frolov-Tseytlin) sectors. 
The beautiful story that followed firmly established the integrable structures in both ${\mathcal{N}}=4$ SYM and IIB string in {\AdSS}  \cite{Frolov:2003xy}-\cite{Arutyunov:2005nk}, and opened up a new fertile testing ground for AdS/CFT correspondence.

To be more detailed, one usually seeks the solitonic solutions to the classical string sigma model on {\AdSS} or certain subspaces of it, and certain rotating string ans\"atze are usually required, the sigma model equations of motion usually simplify into those of classical Neumann or more general Neumann-Rosochatius integrable systems under these ans\"atze \cite{Arutyunov:2003za}. The so-called ``folded'' and ``circular'' solutions of Neumann system are two different types of   
semi-classical strings which extend and rotate in the bulk, their energies and angular momenta can be extracted from elliptic integrals in general \cite{Arutyunov:2003uj,Arutyunov:2003rg}. Sometimes the folded and circular strings are combined to be called the ``elliptic'' solutions to Neumann integrable system.
Remarkably, the folded and circular string solutions, with distinct target space topologies, respectively correspond to the so-called ``double contour'' and ``imaginary'' Bethe roots distributions of the corresponding spin-chain \cite{Beisert:2003xu, Arutyunov:2003uj,Arutyunov:2003rg}! Not only the usual agreement of the string energy with the anomalous dimension, but also the higher conserved charges derived from Neumann integrable system via B\"acklund transform were reproduced from the resolvent of the corresponding spin-chain \cite{Arutyunov:2003rg}.
     
Complementary to the explicit Bethe ans\"atz techniques for spin chains and the construction of rotating strings in the curved background, where one usually make full use of the integrabilities, another line of approach was developed from the work by Kruczenski \cite{Kruczenski:2003gt}.
Here one can compare the spin chain and string sigma model directly at the level of the effective actions, without using the integrabilities. The agreements between the so-called ``coherent state'' action for the spin-chain with the string sigma model action in certain limit would then also imply the match between the particular solutions for the spin chain Bethe equations and the sigma model equations of motion.                       

Understanding the possible correspondence between spin-chain/spinning-string and extend the applications of integrabilities in the less supersymmetric set-up are the main focuses of this paper. While there are many possible SUSY-preserving deformations to ${\mathcal{N}}=4$ SYM, the specific case of our interests is the one developed by Leigh and Strassler \cite{Leigh:1995ep}, the corresponding supergravity dual has recently been discovered in an elegant paper by Lunin and Maldacena \cite{Lunin:2005jy}. What makes Leigh-Strassler deformation special is that it is a continuous, ${\mathcal{N}}=1$ SUSY-preserving, exactly marginal deformation of the maximally supersymmetric theory, the resultant theory is superconformal for all values of gauge coupling, and most importantly, posesses a weakly coupled regime for the perturbative calculations to be allowed. The deformation parameter is usually denoted as $\beta$ which in general can be complex, however, we shall only be concerned with real $\be$ and denote it as $\ga$ instead in this note. 
On the string sigma model side, the simple solution generating techniques introduced in \cite{Lunin:2005jy}, which involves sequence of T-duality transformations and shifts of angular coordinates (subsequently known as ``TsT-transformation'' \cite{Frolov:2005dj}), have produced various new smooth supergravity backgrounds corresponding to multi-parameter deformations of AdS backgrounds \cite{Gauntlett:2005jb}.

The pioneering work in understanding the integrabilities in Leigh-Strassler theory or equivalently Lunin-Maldacena background was done in \cite{Frolov:2005dj,Frolov:2005ty} (For some earlier work, see \cite{Berenstein:2004ys,Roiban:2003dw}), where various techniques used for the maximally supersymmetric theory were extended and generalized. There have subsequently been further interesting related work appearing in literature \cite{Alday:2005ww}-\cite{Mauri:2005pa}. In \cite{Chen:2005sb}, following \cite{Frolov:2005ty}, the authors performed an explicit check of the spinning-string/spin-chain correspondence in Lunin-Maldacena background for particular examples. The precise string energy was reproduced from the double contour solution for the ``twisted'' Bethe ans\"atz equations, this gave the one-loop anomalous dimension for the operator of the form ${\rm Tr}(\Phi_{1}^{J_{1}}\Phi_{2}^{J_{2}})$ and confirmed some independent observations made in \cite{Frolov:2005ty}.
     
In this note, one of the main purposes is to show the spinning string solutions found in \cite{Dimov:2004qv} resulting from the Neumann-Rosochatius system are indeed realized on the Lunin-Maldacena backgrounds, and again provide two kinds of solitonic solutions with two spins.
One of them has been obtained in \cite{Chen:2005sb}, where it was viewed as a perturbation away from the usual two-spin folded string solution in the undeformed background.
However we explore the same class of states/operators as \cite{Frolov:2005ty} in this note and the solutions are not necessarily regarded as perturbations from the undeformed cases.
Here our stance is that the string solutions proposed in this note in general reduce to the ones in \cite{Dimov:2004qv} in the vanishing-$\ga$ limit, and we do not assume them to be directly reduce to the two spin ``circular'' and ``folded'' solutions of the Neumann systems.
We also explore the relationship between different semi-classical string solutions in the Lunin-Maldacena background, as well as the connection between different Bethe root distributions for the corresponding twisted spin chain.

To emphasize more on the correspondence between the spin-chain and spinning string in the deformed cases, we first adopt a different approach from \cite{Chen:2005sb}, instead we will use the Landau-Lifshitz sigma model approach as in \cite{Kruczenski:2003gt,Kruczenski:2004kw}, and show the equivalence between the coherent state action of twisted spin chain and the fast-string limit of sigma model action in Lunin-Maldacena background. This part overlaps with the materials presented in \cite{Frolov:2005ty}, however we decide to keep it for completeness as well as being a good entry route for establishing the correspondence. 

While folded and circular string in the undeformed background are elliptic solutions of the Neumann system, solving the string equations of motion in the deformed background, one usually has to modify the ans\"atz and introduce extra spatial dependence for the $U(1)$ variables, the equations of motion then reduce to those of Neumann-Rosochatius system in general. We show that, for special ratio of winding and oscillation numbers, our semi-classical solutions can be shown to reduce to their Neumann counterparts in the undeformed background. However, in general, our solutions should be relate to those of Neumann-Rosochatius system in the undeformed background as we mentioned earlier. 
Moreover, we establish the relationship between the ``folded'' and ``circular'' strings, demonstrating how an analytic continuation allows us to derive the energy expression for one from the other. 
            
In the twisted spin chain analysis, we propose explicit ans\"atze of Bethe root distributions for each of the two different semi-classical string solutions in the Lunin-Maldacena background, complement the analysis in \cite{Chen:2005sb}. Moreover, the analytic continuation can also be used in deriving the energy for the deformed imaginary distribution from the deformed double contour solutions.        
Once again, we obtain striking match between the deformed circular string energy and the one-loop anomalous dimension calculated from twisted the spin-chain with the deformed imaginary Bethe roots distribution.

This note is organized as follows.
In section \ref{sec:beta-SYM}, we first briefly review the Leigh-Strassler deformation of {\Nf} SYM and the dual Lunin-Maldacena background, the equivalence between the coherent state path integral of the twisted spin-chain Hamiltonian and the fast-string limit of string sigma model is then presented. Section \ref{sec:beta-solutions} is devoted to two explicit semi-classical string solutions with two spins in the Lunin-Maldacena background, using the action derived from section \ref{sec:beta-SYM}. In section \ref{sec:comparison}, we present the detailed twisted spin-chain analysis partly based on the results in \cite{Frolov:2005ty,Chen:2005sb}. We also propose ans\"atz for the outlines of Bethe strings in the scaling limit.
We present our summary and outlook in section \ref{sec:conclusion-discussions}.
In Appendix \ref{app:elliptic}, our conventions for the complete elliptic integrals and some useful integral formulae are listed.
Appendix \ref{sec:RH-beta} contains some key formulae for the twisted spin-chain analysis and sketch the Riemann-Hilbert problem involved.

\section{The Agreement at the Level of Landau-Lifshitz Actions\label{sec:beta-SYM}}
The agreement between the coherent state path integral for the twisted
spin chain associated with the 
Leigh-Strassler theory and 
the effective action for the deformed $SU(2)$ sector of the string theory on
Lunin-Maldacena background was shown in \cite{Frolov:2005ty}, where the calculation was performed
for general complex deformation parameter.
In this section, we begin with a brief review on Leigh-Strassler
theory and Lunin-Maldacena background, followed by an overview of the
aforementioned agreement, the aim here is to make this paper more
self-contained, and highlight the correspondence between spin-chain
and spinning string in the deformed set-up.

\subsection{Brief Review on Deformed Theories and Coherent State Action for Twisted Spin Chain}

It is well-known that $\N=4$ SYM in four dimensions is a
superconformal field theory 
with the complexified coupling constant $\tau=\f{4\pi i}{g_{\rm YM}^{2}}+\f{\theta}{2\pi}$ parameterizing 
a whole family of theories with sixteen supercharges.
The $SL(2,{\mathbb{Z}})$ electric-magnetic duality transformation acts on $\tau$ and relates different theories on the fixed line.   
In addition to $\tau$, Leigh and Strassler considered the further two
$\N=1$ SUSY-preserving, exactly marginal deformations 
of {\Nf} SYM, given by 
${\mathcal{O}}_{1}=h_{1}{\rm Tr}(\Phi_{1}\{\Phi_{2},\Phi_{3}\})$ 
and ${\mathcal{O}}_{2}=h_{2}{\rm Tr}(\Phi_{1}^{3}+\Phi_{2}^{3}+\Phi_{3}^{3})$, 
where $\Phi_{i}$ $(i=1,2,3)$ are three chiral scalars in the
$\N=4$ vector multiplet written in $\N=1$ language. 
The two complex exactly marginal couplings $h_{1}$ and $h_{2}$, along
with $\tau$
now parametrize the whole family of $\N=1$ 
superconformal field theories. 

We are interested in the so-called $\beta$-deformation in this paper,
which corresponds to setting $h_{2}=0$, 
so that up to rescaling, the resultant $\N=1$ superconformal field theory has the superpotential of the form 
\begin{equation}
W_{\be }=\kappa\, \Tr\ko{e^{i\pi\be }\Phi_{1}\Phi_{2}\Phi_{3}-e^{-i\pi\be }\Phi_{1}\Phi_{3}\Phi_{2}}\,,
\end{equation}
with $\kappa$ and $\be$ being complex in general. However, we shall
restrict our attention to real $\be$, 
which we will thereafter denote it as $\gamma$ instead in this paper. 
The $\gamma$-deformation preserves the $U(1)\times U(1)\times U(1)$
Cartan subalgebra 
of the $SU(4)$ R-symmetry of $\N=4$ SYM. 
The linear combination of these three $U(1)$ gives rise to the
$U(1)_{\rm R}$ symmetry and 
the global $U(1)\times U(1)$ symmetry of the resultant $\N=1$ theory. 
One should also note that the $SL(2,{\mathbb{Z}})$ invariance of
$\N=4$ SYM also extends to $\gamma$-deformed theory, 
relating different theories on the fixed plane \cite{Beta-deformed}.  

As we are mainly concerned with the anomalous dimensions for the class
of operators 
of the form $\Tr (\Phi_{1}^{J_{1}}\Phi_{2}^{J_{2}})$ with $J_{1}$, $J_{2}$ and $L\eq J_{1}+J_{2}$ large, 
this particular subsector is known to be closed under one-loop renormalization, we shall therefore focus on the interaction given by
\begin{equation}\label{scalar-pot}
V_{\ga }=\Tr\left|\Phi_{1}\Phi_{2}-e^{-2\pi i \ga }\Phi_{2}\Phi_{1}\right|^{2}\,.
\end{equation} 
For more detailed discussion of the gauge invariant operators in Leigh-Strassler theory, we refer readers to \cite{Frolov:2005ty,Chen:2005sb}.

The supergravity background dual to the $\gamma$-deformation has
recently been found in \cite{Lunin:2005jy}, 
where the $U(1)\times U(1)$ global symmetry was exploited in
generating the new background. 
More specifically, considering the relation between the global
symmetry of $\N=4$ SYM and the resultant $\N=1$ $\gamma$-deformed 
theory, the two torus associated with the $U(1)\times U(1)$ symmetry
should present in the undeformed background {\AdSS} 
and be preserved under the deformation. At the level of supergravity,
the $SL(2,\mathbb{R})$ group associated with the 
two torus acts on the torus parameter, and allows us to generate a
nontrivial NS-NS $B$-field. 
As the result, the background gets deformed by the non-trivial field strength.
The action of the $SL(2,\mathbb{R})$ can also be decomposed into
a sequence of T-duality transformations 
and shift of angular coordinates, referred to as ``TsT-transformation'' in
\cite{Frolov:2005dj}, 
and there has been many applications for this technique in the literature \cite{Gauntlett:2005jb} to generate various deformed backgrounds.

By considering the relevant Feynman diagrams derived from (\ref{scalar-pot}),
the one-loop dilatation operator for the $SU(2)_{\ga}$ sector of the
$\ga$-deformed theory can be shown to be identical 
to the following Hamiltonian of a ferromagnetic XXZ spin chain (without parity invariance) \cite{Frolov:2005ty}:
\begin{equation}
H_{\ga }=\sum_{l=1}^{L}\H_{\ga }^{l,l+1}\,,
\end{equation}
with the nearest-neighbor Hamiltonian density for the link $l$-$(l+1)$,
\begin{align}
\H_{\ga }^{l,l+1}&=\f{\lambda}{16\pi^{2}}\kko{
\ko{{\bf 1}_{l}\otimes {\bf 1}_{l+1}-\sigma^{z}_{l}\otimes \sigma^{z}_{l+1}}
-\komoji{\f{1}{2}}\ko{e^{-2\pi i \ga }\sigma^{+}_{l}\otimes \sigma^{-}_{l+1}+e^{2\pi i \ga }\sigma^{-}_{l}\otimes \sigma^{+}_{l+1}}
}\cr
&=\f{\lambda}{16\pi^{2}}\Big[
\ko{{\bf 1}_{l}\otimes {\bf 1}_{l+1}-\vec\sigma_{l}\otimes \vec\sigma_{l+1}}
+\ko{1-\cos\ko{2\pi\ga }}\ko{\sigma^{x}_{l}\otimes \sigma^{x}_{l+1}+\sigma^{y}_{l}\otimes \sigma^{y}_{l+1}}\cr
&\hspace{2.5cm}{}+\sin\ko{2\pi\ga }\ko{\sigma^{x}_{l}\otimes \sigma^{y}_{l+1}-\sigma^{y}_{l}\otimes \sigma^{x}_{l+1}}
\Big]\,.
\end{align}
Here $\vec\sigma_{l}=\ko{\sigma^{x}_{l},\sigma^{y}_{l},\sigma^{z}_{l}}$ are Pauli matrices at site $l$ and $\sigma^{\pm}_{l}\eq \sigma^{x}_{l}\pm i\sigma^{y}_{l}$.  Turing off $\ga$, we see the spin chain Hamiltonian for the $SU(2)$ sector of the original {\Nf} SYM is recovered:
\begin{equation}
H_{\ga =0}=\f{\lambda}{16\pi^{2}}\sum_{l=1}^{L}\kko{{\bf 1}_{l}\otimes {\bf 1}_{l+1}-\vec \sigma_{l}\otimes \vec \sigma_{l+1}}\,.
\end{equation}

Following \cite{Kruczenski:2003gt, Frolov:2005ty,Fradkin:coherent}, let us perform a so-called ``coherent state path integral'' to obtain an effective action. First we consider the path integral for one spin, i.e., for one site in the chain.
A coherent state $\ket{n}_{l}$ at site $l$ is defined as 
\begin{equation}\label{coh-vec}
\ket{n_{l}}\eq \ket{n(\theta_{l},\phi_{l})}\eq e^{-i\theta_{l}\ko{\sin\phi_{l}\, \sigma^{x}-\cos\phi_{l}\, \sigma^{y}}/2}\ket{0}\,,
\end{equation}
where $\ket{0}$ is the highest weight state of the spin-$\f{1}{2}$ representation.  
The coherent state is defined to have the following remarkable properties: First, the expectation value of Pauli matrices in a coherent state (\ref{coh-vec}) gives an unit three-vector parametrized by $\theta$ and $\phi$, i.e.,
\begin{equation}
\vec n_{l}\eq \bra{n_{l}}\vec\sigma\ket{n_{l}}
=\ko{\sin\theta_{l}\cos\phi_{l},\, \sin\theta_{l}\sin\phi_{l},\, \cos\theta_{l}}\,.
\end{equation}
The ``North Pole'' of the three-sphere would be represented as $\vec n_{0}\eq \ko{0,0,1}$.
Second, the inner product of two coherent states $\ket{n_{1}}$ and $\ket{n_{2}}$ is given by
\begin{equation}
\langle n_{1} | n_{2}\rangle
=\komoji{\ko{\f{1+\vec n_{1}\cdot \vec n_{2}}{2}}^{1/2}} e^{i{\cal A}\ko{\vec n_{1},\vec n_{2},\vec n_{0}}}\,,
\end{equation}
where $\A$ denotes the oriented area of the spherical triangle with vertices at $\vec n_{1}$, $\vec n_{2}$ and $\vec n_{0}$.  When performing the coherent state path integral, the factor containing $\vec n_{1}\cdot \vec n_{2}$ does not contribute to the final expression, and the rest $e^{i{\cal A}\ko{\vec n_{1},\vec n_{2},\vec n_{0}}}$ produces the following so-called ``Wess-Zumino'' term,
\begin{equation}
\SS_{\rm WZ}[\vec n_{l}]=\int {\cal A}_{l}=\int_{0}^{1}d\rho\int dt\, \vec{n}_{l}(t,\rho)\cdot \kko{\pa_{t}\vec{n}_{l}(t,\rho)\times \pa_{\rho}\vec{n}_{l}(t,\rho)}\,.
\end{equation}
Here $\vec{n}(t,\rho)$ $(0\leq \rho \leq 1)$ is an extension of $\vec{n}(t)$ defined such that $\vec{n}(t,0)\eq \vec{n}(t)$
and $\vec{n}(t,1)\eq \vec{n}_{0}$.  Note that, as we are considering a classical solution, the Wess-Zumino term can be partially integrated to give a $\rho$-independent expression like $\pa_{t}\phi \cos\theta$, which turns out to be identical to the undeformed case, i.e., the Wess-Zumino term is not affected by the deformation.

The total action is the sum of the Wess-Zumino term and the expectation value of the $\ga$-deformed Hamiltonian in the coherent states,
\begin{equation}\label{coh-action-def}
\SS [\vec n_{l}] = \f{1}{2}\,\SS_{\rm WZ}[\vec n_{l}]+\int dt \bra{n_{l}} \H_{\ga } \ket{n_{l}}\,.
\end{equation}
In the scaling limit where $\lambda/L^{2}$ and $\ga  L$ are fixed
finite but $L$ is large,
the expectation values in (\ref{coh-action-def}) can be evaluated using
\begin{align}
&\ko{1-\cos\ko{2\pi\gamma}}\bra{n_{l}} \ko{\sigma^{x}_{l}\otimes \sigma^{x}_{l+1}+\sigma^{y}_{l}\otimes \sigma^{y}_{l+1} } \ket{n_{l+1}}
~\sim~ \ko{2\pi\gamma}^{2}\cdot \sin^{2}\theta\,,\cr
&\sin\ko{2\pi\gamma}\bra{n_{l}} \ko{\sigma^{x}_{l}\otimes \sigma^{y}_{l+1}-\sigma^{y}_{l}\otimes \sigma^{x}_{l+1} } \ket{n_{l+1}}
~\sim~ \ko{2\pi\gamma}\cdot \phi' \sin^{2}\theta \cdot \komoji{\f{2\pi}{L}}\,,\cr
&\bra{n_{l}} \ko{{\bf 1}_{l}\otimes {\bf 1}_{l+1}-\vec\sigma_{l}\otimes \vec\sigma_{l+1} } \ket{n_{l+1}}
~\sim~ \komoji{\f{1}{2}}\ko{\theta'{}^{2}+\phi'{}^{2}\sin^{2}\theta} \cdot \komoji{\ko{\f{2\pi}{L}}}^{2}\,,\nonumber
\end{align}
where we defined $\vec n(\hs)=\ko{\sin\theta (\hs)\cos\phi (\hs),\sin\theta (\hs)\sin\phi (\hs),\cos\theta (\hs)}$ with the identification $\vec n(\komoji{\f{2\pi l}{L}})\eq \vec n_{l}$, and the prime $(\, {}'\, )$ denotes a derivative with respect to such defined $\hs$.
Plugging these into (\ref{coh-action-def}), we arrive at the following
effective action,
\begin{align}\label{coh-action-eff}
S_{\rm eff} &=\sum_{l=1}^{L}\SS [\vec n_{l}]
=\f{L}{2\pi}\int_{0}^{2\pi} d\hs\, \SS [\vec n (\hs)]\cr
&= \f{L}{4\pi}\int_{0}^{2\pi} d \hs \int dt\, \pa_{t}\phi \cos\theta
-\f{\lambda}{16\pi L}\int_{0}^{2\pi} d \hs \int dt \kko{\theta'{}^{2}+\ko{\phi'+\ga  J}^{2}\sin^{2}\theta}\,,
\end{align}
which turns out to be an anisotropic Landau-Lifshitz action
 \cite{Frolov:2005ty}. 
We will derive the same expression (\ref{coh-action-eff}) on the string sigma model side in the next subsection.

\subsection{Large-Spin Limit of Spinning Strings on \bmt{\R_{\rm t}\times {\rm S}^{3}_{\ga }}\label{sec:beta-string}}

We shall consider a spinning string solution in the supergravity background dual to the $\ga\, (\in {\mathbb R})$-deformed {\Nf} SYM theory.
Let us first review some relevant aspects of the undeformed case.  
The metric of $\R_{\rm t}\times {\rm S}^{3} \ko{ \subset {\rm AdS}_{5}\times {\rm S}^{5}}$ subspace can be parametrized as
\begin{equation}
ds^{2}_{\R_{\rm t}\times {\rm S}^{3}}=-dt^{2}+\abs{d\xi_{1}}^{2}+\abs{d\xi_{2}}^{2}\,,
\end{equation}
where $t$ is the AdS-time, and the complex coordinates $\xi_{j}$ $(j=1,2)$ are defined by four real embedding coordinates of ${\rm S}^{3}$, $X_{M}$ $(M=1,\dots, 4)$, as $\xi_{1}=X_{1}+i X_{2}$ and $\xi_{2}=X_{3}+i X_{4}$ with $\sum_{M=1}^{4} X_{M}^{2}=\sum_{j=1}^{2}\abs{\xi_{j}}^{2}=1$.
When we consider spinning string solutions, it is useful to introduce the following parametrization with global coordinates, $\xi_{j}=r_{j}e^{i\vp_{j}}$ $(0\leq \vp_{j}< 2\pi)$ with $\sum_{j=1,2} r_{j}^{2}=1$.
Then the Polyakov action for the string which stays at the center of the AdS${}_{5}$ and rotating on the five-sphere takes the form,
\begin{align}\label{action}
S_{\R_{\rm t}\times {\rm S}^{3}}
&=-\f{\sqrt{\lambda}}{2}\int d\tau\int\f{d\sigma}{2\pi}\kkko{
\ga^{\al\be}\kko{-\pa_{\al}t\pa_{\be}t+\pa_{\al}\xi_{j}\pa_{\be}\xi_{j}^{*}+\Lambda\ko{\xi_{j}\xi_{j}^{*}-1}}
}\cr
&=-\f{\sqrt{\lambda}}{2}\int d\tau\int\f{d\sigma}{2\pi}\kkko{
\ga^{\al\be}\kko{-\pa_{\al}t\pa_{\be}t+\pa_{\al}r_{j}\pa_{\be}r_{j}+r_{j}^{2}\pa_{\al}\vp_{j}\pa_{\be}\vp_{j}+\Lambda\ko{r_{i}^{2}-1}}
}
\end{align}
Here $\sqrt{\lambda}=R^{2}/\al'$, with $R$ being the radius of the {\AdSS}, and
$\Lambda$ is a Lagrange multiplier ensuring the sigma model constraint.
We take the standard conformal gauge, $\ga^{\al\be}={\rm diag}(-1,+1)$.
Then the Virasoro constraints are given by
\begin{align}\label{Virasoro}
0=-\ko{\pa_{\tau} t}^{2}+\pa_{\tau}\xi_{j}\pa_{\tau}\xi_{j}^{*}+\pa_{\sigma}\xi_{j}\pa_{\sigma}\xi_{j}^{*}\quad 
\mbox{and}\quad 
0=\pa_{\tau}\xi_{j}\pa_{\sigma}\xi_{j}^{*}\,.
\end{align}

We are interested in the semi-classical string states with two spins
on the deformed ${\rm S}^{3}$ part, 
which we will denote as ${\rm S}^{3}_{\ga }$, of the Lunin-Maldacena
background.  
The trick to generate the supergravity background dual to the
$\ga$-deformed {\Nf} SYM is established in \cite{Lunin:2005jy}, 
known as a ``TsT-transformation''. The recipe is made up of the following three steps: 
(i) Perform a T-duality transformation with respect to one of the $U(1)$ isometries, say $\vp_{1}$.
(ii) Shift another $U(1)$ isometry variable $\vp_{2}$ as $\vp_{2}\to \vp_{2}+\hg\vp_{1}$ with a real parameter $\hg$.
(iii) T-dualize back on $\vp_{1}$.
Applying this TsT-transformation, the deformed background is then given by
\begin{align}\label{action-beta}
S_{\R_{\rm t}\times {\rm S}^{3}_{\ga }}&=-\f{\sqrt{\lambda}}{2}\int d\tau\int\f{d\sigma}{2\pi}\Big\{
\ga^{\al\be}\kko{-\pa_{\al}t\pa_{\be}t+\pa_{\al}r_{j}\pa_{\be}r_{j}+Gr_{j}^{2}\pa_{\al}\vp_{j}\pa_{\be}\vp_{j}+\Lambda\ko{r_{j}^{2}-1}}\cr
&\hspace{4.0cm}{}-2\epsilon^{\al\be}\hg G r_{1}^{2}r_{2}^{2}\pa_{\al}r_{1}\pa_{\be}r_{2}+\Lambda\ko{r_{j}^{2}-1}
\Big\}
\end{align}
with the $\hg$-dependent factor $G=\ko{1+ \hg ^{2}r_{1}^{2}r_{2}^{2}}^{-1}$.
The $\epsilon^{\al\be}$ is the antisymmetric tensor with the signature $\epsilon^{\tau\sigma}=1$, and 
the deformation parameter $\hg$ is related to the parameter $\ga $ in the SYM side as $\hg=\sqrt{\lambda}\, \ga$.
Now let us set $r_{1}=\cos\psi$ and $r_{2}=\sin\psi$ with $0\leq \psi \leq \pi/2$, and define new angle variables by $\zeta\eq\f{\vp_{1}+\vp_{2}}{2}$ and $\eta\eq\f{\vp_{1}-\vp_{2}}{2}$.
In terms of these angles, the three-sphere can be parametrized by $U_{j}e^{i\zeta}$ $(j=1,2)$, where $U_{1}=\cos\psi\,e^{i\eta}$ and $U_{2}=\sin\psi\,e^{-i\eta}$ are ${\mathbb{CP}}^{1}$ coordinates.  
Note that $t$ and $\zeta$ are ``fast'' variables that have no
counterparts in gauge theory side, they should therefore be gauged away
through appropriate constraints 
so that the sigma model action reduces to the one written in terms of only the ``slow'' variables $\psi$ and $\eta$.
The $\ga$-deformed Lagrangian then takes the form
\begin{align}
\L_{\R_{\rm t}\times {\rm S}^{3}_{\ga }}&=-\f{\sqrt{\lambda}}{2}\big\{
\ga^{\al\be}\kko{-\pa_{\al}t\pa_{\be}t+\pa_{\al}\psi\pa_{\be}\psi+G\ko{\pa_{\al}\zeta\pa_{\be}\zeta+\pa_{\al}\eta\pa_{\be}\eta}+2\cos\ko{2\psi}\pa_{\al}\zeta\pa_{\be}\eta}\cr
&\hspace{2.0cm}{}-\hg G \sin^{2}\ko{2\psi}\epsilon^{\al\be}\pa_{\al}\zeta\pa_{\be}\eta
\big\}
\end{align}
As usual, we gauge-fix the AdS-time as $t=\kappa\tau$, which solves the equation of motion $\pa^{2} t=0$.  
We make one more change of variables as $u\eq \zeta-t$ so that $\dot
u$ behaves as $\kappa^{-1}+\ord{\kappa^{-3}}$.  
Then Virasoro constraints are written as
\begin{align}
0&= \kappa^{2}+\dot\psi^{2}+\psi'{}^{2}+G\kko{-\kappa^{2}+\dot u^{2}+ u '{}^{2}\dot\eta^{2}+\eta'{}^{2}+2\kappa\dot u+2\cos\ko{2\psi}\ko{\kappa\dot\eta+\dot u\dot\eta+u'\eta'}}\,,\\
0&= \dot\psi \psi'+G\kko{\dot u u'+\dot\eta \eta'+2\kappa u'+2\cos\ko{2\psi}\ko{\kappa \eta'+\dot u \eta'+\dot\eta u'}}\,.
\end{align}
Here the dots $(\, \dot{} \,)$ and the primes $(\, {}' \,)$ denote the derivatives with respect to the worldsheet time- $(\tau)$ and the space- $(\sigma)$ coordinate.
To get the string solutions whose energy behaves as $E \sim
J+\dots$ in the large-spin limit $J\to \infty$, 
it is needed for rescaling the worldsheet time variable and taking a
special limit, and selecting out the sector we are interested in.  
Following the original paper \cite{Kruczenski:2003gt}, we adopt the following limit:
\begin{equation}\label{Kruczenski_limit}
\kappa\to \infty\,,\quad 
\dot X\to 0\,,\quad 
\kappa \dot X~:~\mbox{fixed}\,,\quad 
X'~:~\mbox{fixed}\qquad 
\mbox{for}\quad 
X=\psi,~u,~\eta\,.
\end{equation}
Note that, when taking the limit (\ref{Kruczenski_limit}) with $\hg  J/\sqrt{\lambda}$ fixed finite, 
the Virasoro constraints (\ref{Virasoro}) become identical to the
ones for the undeformed case.
To match the string action with the effective action of gauge side, it is needed to use these reduced Virasoro constraints and also remove the total derivative term.
Further we should change variables such that $-2\eta\mapsto \phi$ $(0\leq
\phi<2\pi)$ 
and $2\psi\mapsto \theta$ $(0\leq  \theta < \pi)$, and rescale the
worldsheet variables as $\ttau=\tau/\kappa$ 
and $\tsig=\sqrt{\lambda}\,\kappa\sigma/J$, the action finally takes the form (using the relation $\hg=\sqrt{\lambda}\, \ga$),
\begin{align}\label{Lagrangian}
S_{\R_{\rm t}\times {\rm S}^{3}_{\ga }}
=\f{J}{4\pi}\int d\ttau d\tsig\, \dot\phi\cos\theta-\f{\lambda}{16\pi
  J}
\int d\ttau d\tsig\kko{\theta'{}^{2}+\ko{\phi'+\ga  J}^{2}\sin^{2}\theta}\,.
\end{align}
Here we have redefined the notations of dots and primes so that $\, \dot{}\, =\pa_{\ttau}$ and ${}'\, =\pa_{\tsig}$.
This is the same Landau-Lifshitz effective action as we saw in the
gauge theory side, Eq.\,(\ref{coh-action-eff}), 
under the identifications $J\eq L$, $\ttau\eq t$ and $\tsig\eq \hs$.

We should note that the procedures we took in this subsection are,
despite its simplicity, 
not applicable for higher loops in $\lambda$.
Instead we should take the 2d T-dual along $\zeta (\sigma)$ and
introduce the T-dualized field $\widetilde \zeta(\sigma)$, 
then gauge-fix as $\widetilde \zeta(\sigma)=J\sigma/\sqrt{\lambda}$ just as was done in \cite{Frolov:2005ty}.
For more details, see \cite{Kruczenski:2003gt,Kruczenski:2004kw,Frolov:2005ty}.

\section{Elliptic Solutions in \bmt{SU(2)_{\ga }} Sector\label{sec:beta-solutions}}
In this section we present explicit semi-classical string solutions in the Lunin-Maldacena background of the folded and circular types by imposing appropriate ans\"atz on the reduced action obtained from the Landau-Lifshitz approach.
They turn out to coincide precisely with the ones derived from more conventional approach used in \cite{Chen:2005sb}. We shall explain carefully the topologies of the semi-classical strings in Lunin-Maldacena background, compare and contrast them with their undeformed counterparts, and calculate their energy expressions for the comparison with the twisted spin chain analysis.

\subsection{The Topology of Semi-Classical Strings}

To obtain spinning strings moving in the ${\rm S}^{3}_{\ga}$ with large-spin, let us set a rotating string ans\"atz.
The equations of motion deduced from the Lagrangian (\ref{Lagrangian}) are given by
\begin{alignat}{3}
\mbox{for}~\theta\,,\qquad 
0&=\dot\phi\sin\theta+\f{\lambda}{2J^{2}}\kko{\theta''-\sin\theta\cos\theta\ko{\phi'+\ga  J}^{2}}\,,
\label{EOM-theta}\\
\mbox{for}~\phi\,,\qquad 
0&=\dot\theta\sin\theta+\f{\lambda}{2J^{2}}\kko{\ko{\phi'+\ga J}\sin^{2}\theta}'\,.
\label{EOM-eta}
\end{alignat}
To obtain an elliptic solution we are interested in, the following ans\"atz is suitable:
\begin{equation}\label{rot.str.ansatz}
\theta=\theta (\hs)\,,\qquad 
\phi=wt+h(\hs)\,.
\end{equation}
This is the same ans\"atz considered in \cite{Dimov:2004qv} as a generalization of a more popular folded string with no extension in the $\vp_{i}$-directions, and can be found in literatures as ``spherical oscillator'' system, e.g. \cite{Toda:elliptic}.
The presence of the $\hs$-dependent field $h(\hs)$ in (\ref{rot.str.ansatz}) is necessary to ensure the equations of motion for $\vp_{i}$ is satisfied. 
Under the (\ref{rot.str.ansatz}), the equation of motion (\ref{EOM-eta}) implies $\ko{\phi'+\ga J}\sin^{2}\theta$ is independent of worldsheet coordinates, and thus we can set that
\begin{align}\label{def-A}
A\eq \ko{\phi'+\ga J}\sin^{2}\theta\quad (\mbox{const.})\,.
\end{align}
Plugging this definition, the equation of motion (\ref{EOM-theta}) reduces to
\begin{align}
\theta''=A^{2}\f{\cos\theta}{\sin^{3}\theta}-\f{1}{2}B\sin\theta\qquad 
\mbox{with}\quad B\eq \f{16\pi^{2}w}{\lambda}\,,
\end{align}
here we can easily integrate this and choose the integration constant $C$ such that
\begin{align}\label{constraint-theta}
\theta'{}^{2}=C+B\cos\theta-\f{A^{2}}{\sin^{2}\theta}\,.
\end{align}
This is the equation that governs the motion of the semi-classical strings, notice that the familiar elliptic solutions of the Neumann system corresponds to case of vanishing $A$.

\begin{figure}[htb]
\begin{center}
\vspace{.7cm}
\hspace{-.0cm}\includegraphics[scale=0.75]{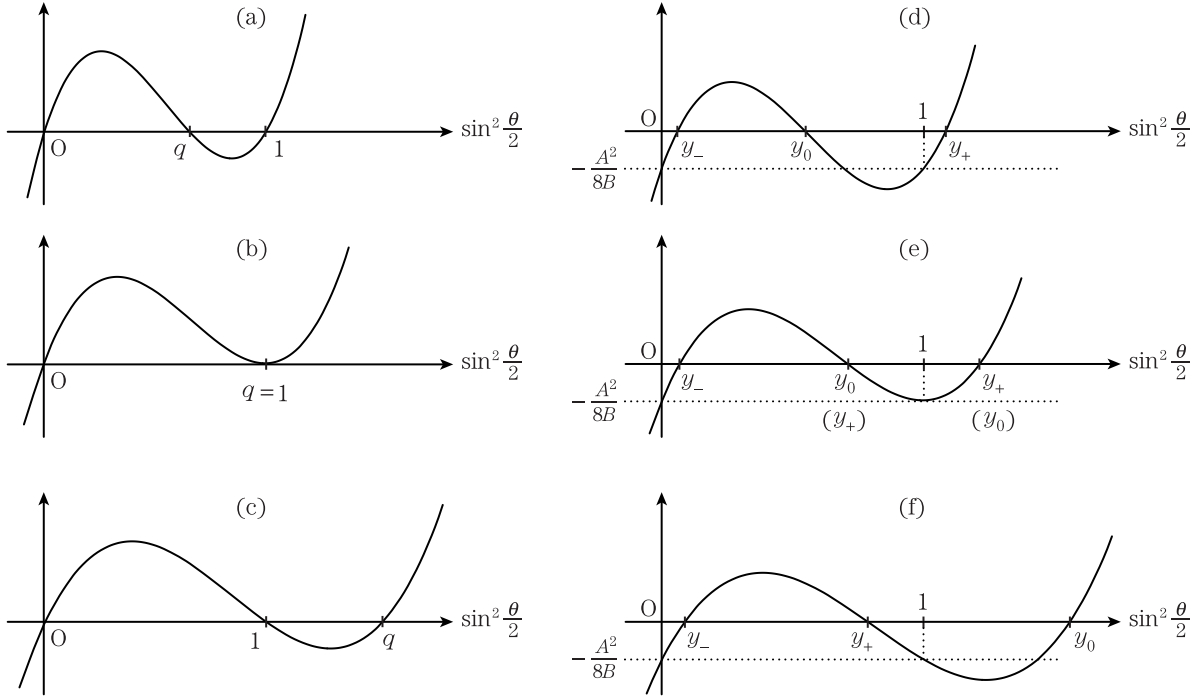}
\vspace{.5cm}
\caption{\small{The graphs of the profile function $f(y)$ in the undeformed (a)-(c) and the $\ga$-deformed cases (d)-(f).
The graph (a) represents a folded string with a moduli $q=\sin^{2}\f{\theta_{0}}2$, folded onto the interval $[-\theta_{0}/2,\theta_{0}/2]$ in the $\theta$- (or $\psi$-)direction. 
The graph (c) represents the circular string, and (b) is the critical point where the transition of folded/circular takes place.
In the vanishing-$\ga$ limit, (d), (e) and (f) reduce to (a), (b) and (c), respectively.}}
\label{fig:transition}
\end{center}
\end{figure}

To proceed, it is convenient to introduce a new parameter
\begin{equation}
y\eq \sin^{2}\f{\theta}{2}\,,\;\;\;\;\;0\leq y\leq 1\,,
\end{equation}
so that the governing equation (\ref{constraint-theta}) can be re-written into the following form
\begin{equation}\label{y'2}
y'{}^{2}=2B\ko{y_{+}-y}\ko{y_{0}-y}\ko{y-y_{-}}\eq 2Bf(y)\,.
\end{equation}
Here $y_{0}$ and $y_{\pm}$ are the three roots of $f(y)=0$ defined such that 
\begin{align}\label{elliptic-moduli-LLSM}
y_{+}+y_{0}+y_{-}=\f{C+3B}{2B}\,,\quad 
y_{+}y_{0}+y_{0}y_{-}+y_{-}y_{+}=\f{C+B}{2B}\,,\quad 
y_{+}y_{0}y_{-}=\f{A^{2}}{8B}\,.
\end{align}
Notice that equations in (\ref{elliptic-moduli-LLSM}) are invariant under the permutations of $y_{+}$, $y_{0}$ and $y_{-}$, such symmetry also appears in spin chain analysis as the freedom in exchanging the end points of the Bethe strings. However, for the right hand side of (\ref{y'2}) to remain non-negative, we are only allowed to exchange roles of $y_{+}$ and $y_{0}$ and the two different choices result in two different topologies for the semi-classical strings in the target space.  
To make this clearer, let us define $y_{-}$ to be the smallest positive root of $f(y)=0$, and $y_{+}$ such that the sign of $\left.\f{df}{dy}\right|_{y_{+}}$ coincides with that of $\left.\f{df}{dy}\right|_{1}$;
the remaining root is identified with $y_{0}$.
The graphs of the function $f(y)$ in various situations are depicted in Fig.\,\ref{fig:transition}.
Observing the profile function $f(y)$ is a degree three curve with $f(0)=f(1)=-\f{A^{2}}{8B}\leq 0$, it follows that only $y_{-}$ and one of the two other roots lie in the physical range range $[0,1]$, while the third root is greater than one.
When $\left.\f{df}{dy}\right|_{1}>0$, $f(y)$ is positive for $y_{-}<y<y_{0}$ (Fig.\,\ref{fig:transition} (d)), whereas when $\left.\f{df}{dy}\right|_{1}<0$, $f(y)$ is positive for $y_{-}<y<y_{+}$ (Fig.\,\ref{fig:transition} (f)).

To visualize the semi-classical strings corresponding to these two different cases, let us first consider the one with ans\"atz $\theta=\theta (\hs)$ and $\phi=wt$. Such ans\"atz corresponds to setting $A=0$ so that $y_{-}=0$, and the first two equations in (\ref{elliptic-moduli-LLSM}) give $y_{+}=1$ and $y_{0}=q\eq\frac{C+B}{2B}$.
Fig.\,\ref{fig:transition} (d) and Fig.\,\ref{fig:transition} (f) reduce in this limit to Fig.\,\ref{fig:transition} (a) and Fig.\,\ref{fig:transition} (c) respectively.
The topology of the semi-classical string is determined by whether $C+B\cos\theta=0$ can be satisfied by any $\theta$ or not, or equivalently whether $q\leq 1$ or $q>1$.  
The two cases are known to be associated with the ``folded'' $(q\leq 1)$ and ``circular'' $(q>1)$ solutions of the Neumann integrable system.
The half-period of the folded case starts from $\theta=0$ and goes to $\theta=\theta_{0}$, then back to $\theta=0$; while in the circular case, the sign of $\theta'$ never changes, $\theta$ can take any values from $0$ to $\pi$. Instead of folding back to itself, the string completely winds one of the great circles of {\S} passing through $\theta=0$. 
One is actually allowed to generalize the ans\"atz for the semi-classical string to have $\phi (t,\hs)=wt+h(\hs)$, such that the string acquires winding profile in the $\phi$-direction, giving a solution associated with so-called Neumann-Rosochatius (NR) integrable system, which was investigated in \cite{Dimov:2004qv} in both by `conventional' and the Landau-Lifshitz approaches.

By contrast, in the deformed background, it is \textit{necessary} having the ans\"atz for the semi-classical string to be $\phi(t,\hs)=wt+h(\hs)$ as in (\ref{rot.str.ansatz}) so that equations of motion can be satisfied. Here the string never reaches $\theta=0$ nor $\theta=\pi$ due to the presence of nonzero $A^{2}$ in (\ref{constraint-theta}). The solutions should not be regarded as only a naive perturbation for the usual solution of the Neumann system;
instead, they should in general be treated as a perturbative solution from that of NR integrable system. 
Here we can still distinguish two different classes of solutions, depending whether $\left.\f{df}{dy}\right|_{1}$ is positive or negative, or equivalently, $y_{0}<1$ or $y_{0}>1$.
Recall that the moduli parameter $q\eq \f{C+B}{2B}$ controls the topology of a classical solution of the Neumann system, then it follows that
\begin{eqnarray}
\left.\f{df}{dy}\right|_{y=1}
\Bigg\{ \begin{array}{ll}
{}\leq 0 & \iff \, 0\leq q\leq 1 \\
{}>0 & \iff \, 1<q \\
\end{array} \,,
\end{eqnarray}
these turn out to be exactly the conditions for a string of Neumann system to be of folded (upper) or of circular (lower) type.
Therefore, despite slight abuse of the terminology, we shall borrow from the Neumann situation, and call the case $y_{0}<1$ ``folded'' string and the case $y_{0}>1$ ``circular'' string even in the deformed background with the integrability of the NR type.

\subsection{``Folded'' Case}
Now we turn to the explicit semi-classical string solutions, and let us begin with the folded case.
The conserved sigma model charges such as the $z$-component of the spin $S_{z}$, total spin $J$ and energy $E$ are calculated as
\begin{align}
S_{z}&=\f{J}{4\pi}\int_{0}^{2\pi} d\hs\int_{0}^{1}d\rho\, \pa_{\rho}\theta\sin\theta
=\f{J}{4\pi}\int^{2\pi}_{0} d\hs\, \cos\theta\,,\\
J&=\f{J}{2\pi}\int_{0}^{2\pi} d\hs\,,\\
E&=\f{\lambda}{16\pi J}\int_{0}^{2\pi} d\hs\ko{\theta'{}^{2}+\ko{\phi'+\ga  J}^{2}\sin^{2}\theta}
=\f{\lambda}{16\pi J}\int_{0}^{2\pi} d\hs\ko{C+B\cos\theta}\,,
\end{align}
respectively.  
We used the definition of $A$ and the constraint (\ref{constraint-theta}) to put the Hamiltonian into the form above.
Let us denote the number of windings along the $\phi$-direction by $N$, and the number of the oscillations in the $\theta$-direction by $M$, that is the number of times the semi-classical string traces out the allowed range of $\theta$.
Performing the integration for $\phi'$ using the formula (\ref{ellip-int-formula-3}), we have
\begin{equation}\label{M-M}
2\pi N
=\int_{0}^{2\pi}d\hs \, \phi'
=-2\pi\ga J+M\sqrt{\f{{y_{+}y_{0}y_{-}}}{y_{+}-y_{-}}}\kko{\f{1}{y_{-}}\PP{\f{y_{0}-y_{-}}{-y_{-}},q}
+\f{1}{1-y_{-}}\PP{\f{y_{0}-y_{-}}{1-y_{-}},q}}\,,
\end{equation}
where we defined the elliptic moduli $q$ as
\begin{equation}\label{moduli-string}
q\eq \f{y_{0}-y_{-}}{y_{+}-y_{-}}\,.
\end{equation}
The Eq.\,(\ref{M-M}) relates the parameters $y_{\pm}$, $y_{0}$ and the deformation parameter $\ga J$. Here $N$ and $M$ need to be integers, as the ans\"atz describes a physical closed string on the smooth deformed background.
From the viewpoint of the gauge theory, Eq.\,(\ref{M-M}) corresponds to a so-called ``trace condition'' (or ``cyclicity condition'') of an associated SYM operator with now a nonzero twist parameter $\ga J$.
If we insist that the solution interpolates to the class of semi-classical string solutions in undeformed background, with $y_{+}$, $y_{-}$ and $y_{0}$ being 1, 0 and some finite value between $0$ and $1$, in such limit, the equation (\ref{M-M}) reduces to $M/N=2$.
In other words, for a special case of our solution to reduce the popular folded strings of Neumann system, we have to start with $M/N=2$, then take the limit $\gamma\to 0$. 
In fact, up to this point, our discussion is generally applicable to either folded or circular string in the deformed background. The same criterion $M/N=2$ is needed if our solution is to interpolate to the popular (elliptic) circular string in the Neumann system with now $y_{0}>1$.

\begin{figure}[htb]
\begin{center}
\vspace{.7cm}
\hspace{-.0cm}\includegraphics[scale=0.55]{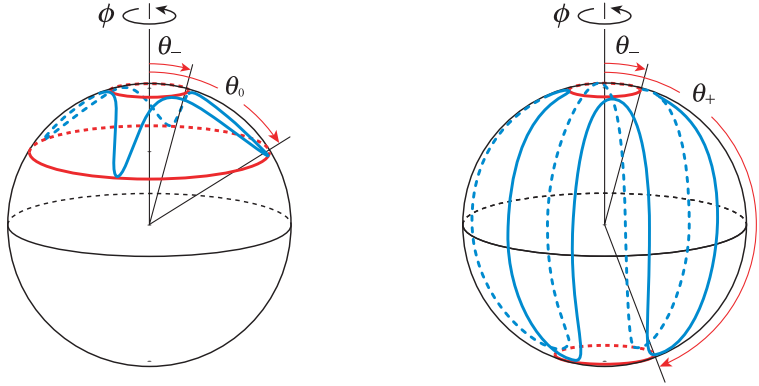}
\caption{\small{Profiles of spinning string solutions in  NR-system with $M/N=4$, projected onto an ${\rm S}^{2}$ of ${\mathbb R}_{\rm t}\times {\rm S}^{3}$. The angles $\theta_{0}$ and $\theta_{\pm}$ are defined such that $\sin^{2}\f{\theta_{0}}{2}=y_{0}$ and $\sin^{2}\f{\theta_{\pm}}{2}=y_{\pm}$.  The left and the right figures are called ``folded'' and ``circular'' type in the text, and correspond to, respectively, (d) and (f) of Fig.\,\ref{fig:transition}.}}
\label{fig:string-on-LM-nutation}
\end{center}
\end{figure}

In more general cases, however, we can have various eccentric versions of folded strings of NR system without constraint on the integers $M$ and $N$.
As an example, see the left figure of Fig.\,\ref{fig:string-on-LM-nutation} for the $M/N=4$ case, whose profile looks just like the trajectories of a precessing top with a nutation.
The right side of Fig.\,\ref{fig:string-on-LM-nutation} represents the eccentric version of a (elliptic) circular string, which will be discussed later.

Let us now focus on $N=1$ case, but keeping $M$ arbitrary.
The total spin can be calculated using the integral formula (\ref{ellip-int-formula-1}),
\begin{align}\label{J}
J=\f{J}{2\pi}\int_{y_{-}}^{y_{0}}\f{2Mdy}{\sqrt{2B\ko{y_{+}-y}\ko{y_{0}-y}\ko{y-y_{-}}}}
=\f{JM}{\pi}\sqrt{\f{2}{B}}\f{\K{q}}{\sqrt{y_{+}-y_{-}}}\,.
\end{align}
Similarly, the $z$-component of the spin $S_{z}$ and the Hamiltonian are given by
\begin{align}
S_{z}&=\f{J}{2}-\f{MJ}{\pi}\sqrt{\f{2}{B}}\f{1}{\sqrt{y_{+}-y_{-}}}\kko{y_{+}\K{q}-\ko{y_{+}-y_{-}}\E{q}}\,,\\
E&=\f{\lambda}{16\pi J}\kkko{4\pi B\kko{y_{+}+2y_{-}-1+q\ko{y_{+}-y_{-}}}-\f{4M\sqrt{2B}}{\sqrt{y_{+}-y_{-}}}\kko{y_{+}\K{q}-\ko{y_{+}-y_{-}}\E{q}}}\,,
\end{align}
where we used the integral formula (\ref{ellip-int-formula-2}).
We can eliminate $B$ by using (\ref{J}).  
Taking into account that the ratio of $J_{2}$ to the total spin $J=J_{1}+J_{2}$ which we denote $\al$ as usual, is related to the sum of the third component of the spin as $S_{z}/J=\f{1}{2}-\al$, the final expression of $\al$ and $E$ for this folded case are then given by
\begin{align}
\al_{\rm fold}&=y_{+}-\ko{y_{+}-y_{-}}\,\f{\E{q}}{\K{q}}\,,\label{ROS-f-LL}\\
E_{\rm fold}&=\f{M^{2}\lambda}{2\pi^{2}J}\,\K{q}
\kko{\E{q}+\ko{(q-1)+\f{y_{+}+y_{-}-1}{y_{+}-y_{-}}}\,\K{q}}\,.\label{ENE-f-LL}
\end{align}
In the vanishing-$\ga$ limit where $y_{+}\to 1$, $y_{-}\to 0$ and $y_{0}\to q$, we see that (\ref{ROS-f-LL}) and (\ref{ENE-f-LL}) indeed recover the expressions for the undeformed case with $M=2$ in \cite{Beisert:2003ea}, see Fig.\,\ref{fig:transition}.

Here we shall summarize the counting of free parameters and the constraints of the system.
At first we had three parameters $y_{\pm}$ and $y_{0}$ to describe the system, which were three roots of $f(y)=0$.
But since there was one condition $f(0)=f(1)$, the number of free parameters was actually two.
One of the degrees of freedom was fixed by giving a set of integers $(M,N)$ and the deformation parameter $\ga J$ through the condition (\ref{M-M}), and the other was done by giving the spin-fraction $\al_{\rm fold}$.
Then we could write down the energy $E_{\rm fold}$ in terms of
so-determined set of moduli parameters.

Let us see how our parametrization of the solution can be mapped to the one used in \cite{Chen:2005sb} where the folded solution was described in the conventional way, set the following rotating string ans\"atz in the original sigma model action (\ref{action-beta}),
\begin{equation}
t=\kappa\tau\,,\quad 
\psi=\psi (\sigma)\,,\quad 
\vp_{i}=w_{i}\tau+h_{i}(\sigma)\,.
\end{equation}
This is the same ans\"atz as introduced in \cite{Arutyunov:2003uj,Dimov:2004qv}, which is known to give us a solution for the NR system.
The $\sigma$-components of the worldsheet currents are now constants and can be written as
\begin{equation}
C_{1}\eq \cJ_{1}^{\sigma}=-r_{1}^{2}G\ko{h_{1}'+\hg  r_{2}^{2}w_{2}}\,,\qquad 
C_{2}\eq \cJ_{2}^{\sigma}=-r_{2}^{2}G\ko{h_{2}'-\hg  r_{1}^{2}w_{1}}\,,
\end{equation}
and it is convenient to define effective angular velocities in the deformed background as $\Omega_{1}\eq w_{2}-\hg  C_{1}$ and $\Omega_{2}\eq w_{1}+\hg  C_{2}$.
In terms of these variables, the differential equation which governs the system can be cast into the following form:
\begin{equation}
\psi'{}^{2}=D-\Omega_{21}^{2}\sin^{2}\psi-\f{C_{1}^{2}}{\cos^{2}\psi}-\f{C_{2}^{2}}{\sin^{2}\psi}\,,
\end{equation}
where $D$ is an arbitrary integration constant and we have defined $\Omega_{21}^{2}\eq \Omega_{2}^{2}-\Omega_{1}^{2}$.
The Virasoro constraints for the system now takes the forms
\begin{align}
\kappa^{2}&=D+\Omega_{1}^{2}+\hg ^{2}\ko{C_{1}^{2}\sin^{2}\psi+C_{2}^{2}\cos^{2}\psi
+\Omega_{1}^{2}\cos^{4}\psi\sin^{2}\psi+\Omega_{2}^{2}\cos^{2}\psi\sin^{4}\psi}\,,\label{Virasoro-1}\\
0&=\Omega_{1}C_{1}+\Omega_{2}C_{2}\,.\label{Virasoro-2}
\end{align}
Note that the latter condition (\ref{Virasoro-2}) reduces to a trivial one in the vanishing-$\ga$ limit, but when we have nonzero $\ga$, it turns out to play a key role in realizing the correspondence between the string and the gauge sides.
The classical energy and the spin-fraction can be described through three parameters $\xz$ and ${x}_{\pm}$ defined such that 
\begin{equation}\label{elliptic-moduli-BA}
\xp+\xz+\xm=1+\f{D}{\Omega_{21}^{2}}\,,\quad 
\xp\xz+\xz\xm+\xm\xp=\f{C_{2}^{2}-C_{1}^{2}+D}{\Omega_{21}^{2}}\,,\quad 
\xp\xz\xm=\f{C_{2}^{2}}{\Omega_{21}^{2}}\,.
\end{equation}
Instead of using $\xz$ and ${x}_{\pm}$, it is more convenient to define a new moduli parameter by
\begin{equation}
k = \f{\xz-\xm}{\xp-\xm}\,,
\end{equation}
and we will describe the system in terms of ${x}_{\pm}$ and $\k$.
To obtain a solution to the first order in the large-spin limit, let us expand all the parameters as $\k=\k^{(0)}+\ord{\f{\lambda}{J^{2}}}$, ${x}_{\pm}={x}_{\pm}^{(0)}+\ord{\f{\lambda}{J^{2}}}$ and $C_{i}=C_{i}^{(0)}+\ord{\f{\lambda}{J^{2}}}$, in which case the Virasoro constraint (\ref{Virasoro-2}) implies $C_{1}^{(0)}+C_{2}^{(0)}=0$.
Setting $\chi\eq C_{1}^{(0)}=-C_{2}^{(0)}$, we see the
relations (\ref{elliptic-moduli-BA}) then reduce to 
\begin{equation}\label{elliptic-moduli-BA-2}
\xp^{(0)}+\xz^{(0)}+\xm^{(0)}=1+\f{D}{\Omega_{21}^{(0)}{}^{2}}\,,\quad 
\xp^{(0)}\xz^{(0)}+\xz^{(0)}\xm^{(0)}+\xm^{(0)}\xp^{(0)}=\f{D}{\Omega_{21}^{(0)}{}^{2}}\,,\quad 
\xp^{(0)}\xz^{(0)}\xm^{(0)}=\f{\chi^{2}}{\Omega_{21}^{(0)}{}^{2}}\,,
\end{equation}
with $\Omega_{21}^{(0)}=\f{2}{\pi}\K{\k^{(0)}}\sqrt{\xp^{(0)}-\xm^{(0)}}$.
Here we see the two elliptic systems derived from two distinct approaches, (\ref{elliptic-moduli-LLSM}) and (\ref{elliptic-moduli-BA-2}), can be identified with each other, under the following identifications of the parameters:
\begin{equation}\label{identification}
B\eq 2\Omega_{21}^{(0)}{}^{2}\,,\qquad 
C+B\eq 4D\,,\qquad 
A^{2}\eq 16\chi^{2}\,.
\end{equation}
The identification (\ref{identification}) implies further mapping of the parameters, ${x}_{\pm}^{(0)}\eq y_{\pm}$ and $k^{(0)}\eq q$, they allow us to rewrite (\ref{ROS-f-LL}) and (\ref{ENE-f-LL}) in terms of $x_{\pm}^{(0)}$ and $k^{(0)}$. Setting $M=2$, the final expressions precisely reproduce Eqs.\,(4.32) and (4.30) in \cite{Chen:2005sb},
\begin{align}
\widetilde \al_{\rm fold}&=\xp^{(0)}-\big( \xp^{(0)}-\xm^{(0)}\big)\f{\E{\k^{(0)}}}{\K{\k^{(0)}}}\,,\label{ROS-f}\\
\epsilon_{\rm fold}^{(1)}&=\f{2}{\pi^{2}}\,\K{\k^{(0)}}\kko{\E{\k^{(0)}}-\big(1-\k^{(0)}\big)\K{\k^{(0)}}+\f{\xp^{(0)}+\xm^{(0)}-1}{\xp^{(0)}-\xm^{(0)}}\,\K{\k^{(0)}}}\,,\label{ENE-f}
\end{align}
where $\widetilde \alpha_{\rm fold}$ is the filling ratio and $\epsilon_{\rm fold}^{(1)}$ is the coefficient of the order $\f{\lambda}{J}$-term in the power series expansion of the energy.

\subsection{``Circular'' Case}
An elliptic circular string solution can be obtained in a similar way as in the folded case. Consider again the $N=1$ case.
For the circular case we have a moduli parameter $y_{0}$ greater than $y_{+}$, and the only difference from the folded case is the range of integration,  we have to change from $\int_{y_{-}}^{y_{0}}$ to $\int_{y_{-}}^{y_{+}}$.
The total spin is given by
\begin{align}
J&=\f{J}{2\pi}\int_{y_{-}}^{y_{+}}\f{2Mdy}{\sqrt{2B\ko{y_{+}-y}\ko{y_{0}-y}\ko{y-y_{-}}}}
=\f{JM}{\pi}\sqrt{\f{2}{B}}\f{\K{1/q}}{\sqrt{y_{0}-y_{-}}}\,.
\end{align}
We can use the condition above to express $B$ in terms of other parameters and give us 
\begin{align}
\al_{\rm circ}&=y_{0}-\ko{y_{0}-y_{-}}\f{\E{1/q}}{\K{1/q}}\,,\label{ROS-c-LL}\\
E_{\rm circ}&=\f{M^{2}\lambda}{2\pi^{2}J}\, \K{1/q}\kko{\E{1/q}+\f{y_{+}+y_{-}-1}{q\ko{y_{+}-y_{-}}}\, \K{1/q}}\label{ENE-c-LL}\,,
\end{align}
for the spin-fraction and the energy of the circular string.
We can also derive the same quantities using the power series expansions,
this is done in much the similar way as in the folded case, and we only need to care the range of where the integration constant $D$ sits in.
The expression for the energy and the spin-fraction are given by, 
\begin{align}
\widetilde \al_{\rm circ}&=\xz^{(0)}-\big( \xz^{(0)}-\xm^{(0)}\big)\kko{\f{\E{1/\k^{(0)}}}{\K{1/\k^{(0)}}}}\,.\label{ROS-c}\\
\epsilon_{\rm circ}^{(1)}&=\f{2}{\pi^{2}}\,\K{1/\k^{(0)}}\kko{\E{1/\k^{(0)}}+\f{(\xp^{(0)}+\xm^{(0)}-1)}
{k^{(0)}\big( \xp^{(0)}-\xm^{(0)}\big)}\,\K{1/\k^{(0)}}}\,.\label{ENE-c}
\end{align}
Comparing these with the Landau-Lifshitz approach, again they coincide with the same identifications as in the folded string case, and the counting of free parameters and the constraints is just the same as the folded case.
Turing off $A$, the known expressions for the circular solution of the Neumann system is recovered. 
Indeed, the reader can check the ``folded'' and the ``circular'' solutions in the deformed background posses a interesting property, they are related to each other via an analytic continuation with respect to the elliptic moduli $q$, just as was presented in \cite{Okamura:2005cj} for the undeformed case.

Finally, taking the rational limit in the elliptic circular case amounts to sending $q$ to $\infty$, we have $E\to \f{M^{2}\lambda}{8J}$,
which corresponds to the ``half-filling'' limit $(\alpha_{\rm circ}\to 1/2)$ of the (rational) circular solution calculated in \cite{Frolov:2005ty}.

\section{Comparison with Twisted Spin-Chain Analysis\label{sec:comparison}}

In this section we shall discuss the gauge theory dual to the semi-classical strings in the Lunin-Maldacena background, and explicitly perform the twisted spin chain analysis.

We first present our results for the one-loop anomalous dimensions calculated from the Bethe ans\"atz, demonstrate how the idea of an analytic continuation can be used in deriving the results for the circular string, based on the existing double contour calculation associated with the folded string in \cite{Chen:2005sb}. As in the undeformed case, the double contour solution in \cite{Chen:2005sb} is strictly valid for even number of the Bethe roots, that is, even number of impurities, while the circular string case should be associated with odd number of Bethe roots.

To complement our results on the string sigma model side and complete
the analysis of the duality for the $\ga$-deformed theories, we will
present a detailed discussion on the distributions of Bethe roots, 
explicitly propose the ans\"atze, from which we can calculate the one-loop anomalous dimensions for the operators associated with the semi-classical string in deformed background.
Remarkably, the parameters describing the endpoints of the Bethe strings will have nice interpretations in terms of the string moduli parameters.

In order to make this note self-contained, in a separate appendix, we shall collect some key formulae for the twisted spin chain analysis and outline how one can obtain the associated Riemann-Hilbert problem in certain scaling limit.


\subsection{``Double Contour'' Case}

Let us first discuss the ``double contour'' case.
In \cite{Chen:2005sb}, a general double contour solution for the twisted spin chain was presented, where the two cuts are located at $\C_{+}\eq [x_{1}, x_{2}]$ and $\C_{-}\eq [x_{3}, x_{4}]$ and the solution was shown to reproduce the energy of the folded string solution in the Lunin-Maldacena background. This should be regarded as a $\gamma$-deformed version of the usual double contour solution. Here we briefly review the key results.
The $\A$- and the $\B$-cycle conditions (\ref{periods}) for the deformed double contour case become as follows:
\begin{align}\label{cycle-cond-DC}
0&=\oint_{\A}dp=2i\int_{x_{2}}^{x_{1}}\f{F(x) dx}{\sqrt{\ko{x_{1}-x}\ko{x-x_{2}}\ko{x-x_{3}}\ko{x-x_{4}}}}\,,\\
4\pi n &=\oint_{\B}dp=2\int_{x_{2}}^{x_{3}}\f{F(x) dx}{\sqrt{\ko{x_{1}-x}\ko{x_{2}-x}\ko{x-x_{3}}\ko{x-x_{4}}}}\,,
\end{align}
where $F(x)$ is a rational function.
The resulting expression for the filling-fraction and one-loop anomalous dimension are (here `DC' stands for `double contour')
\begin{align}
\al_{\rm DC}&=\f{1}{2}-\f{x_{1}x_{2}+x_{3}x_{4}}{4\sqrt{x_{1}x_{2}x_{3}x_{4}}}
+\f{\ko{x_{1}-x_{3}}\ko{x_{2}-x_{4}}}{4\sqrt{x_{1}x_{2}x_{3}x_{4}}}
\f{\E{\f{\ko{x_{1}-x_{2}}\ko{x_{3}-x_{4}}}{\ko{x_{1}-x_{3}}\ko{x_{2}-x_{4}}}}}{\K{\f{\ko{x_{1}-x_{2}}\ko{x_{3}-x_{4}}}{\ko{x_{1}-x_{3}}\ko{x_{2}-x_{4}}}}}\nonumber\\
&=\f{1}{2}-\f{x_{1}x_{2}+x_{3}x_{4}}{4\sqrt{x_{1}x_{2}x_{3}x_{4}}}
+\f{\ko{x_{1}-x_{4}}\ko{x_{2}-x_{3}}}{4\sqrt{x_{1}x_{2}x_{3}x_{4}}}
\f{\E{\f{\ko{x_{1}-x_{2}}\ko{x_{3}-x_{4}}}{\ko{x_{2}-x_{3}}\ko{x_{4}-x_{1}}}}}{\K{\f{\ko{x_{1}-x_{2}}\ko{x_{3}-x_{4}}}{\ko{x_{2}-x_{3}}\ko{x_{4}-x_{1}}}}}\,,
\label{fraction-DC}\\
\gamma_{\rm DC}&=\f{n^{2}}{32 \pi^{2}}\ko{\f{1}{x_{2}}-\f{1}{x_{4}}}\ko{\f{1}{x_{1}}-\f{1}{x_{3}}}\,
\f{\E{\f{\ko{x_{1}-x_{2}}\ko{x_{3}-x_{4}}}{\ko{x_{1}-x_{3}}\ko{x_{2}-x_{4}}}}}{\K{\f{\ko{x_{1}-x_{2}}\ko{x_{3}-x_{4}}}{\ko{x_{1}-x_{3}}\ko{x_{2}-x_{4}}}}}
-\f{n^{2}}{128\pi^{2}}\kko{\ko{\f{1}{x_{1}}+\f{1}{x_{2}}}-\ko{\f{1}{x_{3}}+\f{1}{x_{4}}}}^{2}\,.
\label{AD-DC}
\end{align}
Along with the idea held in \cite{Chen:2005sb} and the identifications between $\{ y_{\pm}, y_{0},q \}$ and $\{ x_{\pm}^{(0)}, x_{0}^{(0)}, k^{(0)} \}$, we can now have the following natural identifications relating the moduli on both spin-chain and string sides in order to realize the duality, which follows from comparing (\ref{ROS-f-LL}) and (\ref{fraction-DC}):
\begin{alignat}{4}\label{dictionary}
&y_{+}=\f{1}{2}-\f{x_{1}x_{2}+x_{3}x_{4}}{4\sqrt{x_{1}x_{2}x_{3}x_{4}}}\,,&\qquad &y_{-}=\f{1}{2}-\f{x_{1}x_{3}+x_{2}x_{4}}{4\sqrt{x_{1}x_{2}x_{3}x_{4}}}\,,\no\\
&y_{0}=\f{1}{2}-\f{x_{1}x_{4}+x_{2}x_{3}}{4\sqrt{x_{1}x_{2}x_{3}x_{4}}}\,,&\qquad &q=-\f{\ko{x_{1}-x_{2}}\ko{x_{3}-x_{4}}}{\ko{x_{1}-x_{4}}\ko{x_{2}-x_{3}}}\,.
\end{alignat}
Using these identifications, we can rewrite the anomalous dimensions (\ref{AD-DC}) and the string energies (\ref{ENE-f-LL}) as
\begin{equation}
\gamma_{\rm DC}=\f{2n^{2}}{\pi^{2}}\,\K{q}\kko{\E{q}-\ko{1-q} \K{q}}-\f{n^{2}}{128\pi^{2}}\kko{\ko{\f{1}{x_{1}}+\f{1}{x_{4}}}-\ko{\f{1}{x_{2}}+\f{1}{x_{3}}}}^{2}\,,
\end{equation}
and
\begin{equation}
E_{\rm fold}=\f{M^{2}}{2\pi^{2}}\,\K{q}\kko{\E{q}-\ko{1-q}\K{q}}
+\f{M^{2}}{128\pi^{2}}\ko{\f{1}{x_{1}}+\f{1}{x_{4}}}\ko{\f{1}{x_{2}}+\f{1}{x_{3}}}\,.
\end{equation}
For the two quantities above to be matched, in addition to $n=M/2$, the following condition is required: 
\begin{align}
&-\kko{\ko{\f{1}{x_{1}}+\f{1}{x_{4}}}-\ko{\f{1}{x_{2}}+\f{1}{x_{3}}}}^{2}
=4\ko{\f{1}{x_{1}}+\f{1}{x_{4}}}\ko{\f{1}{x_{2}}+\f{1}{x_{3}}}\cr
&\quad \Rightarrow\quad 
\sum_{k=1}^{4}\f{1}{x_{k}}=0\,,\label{sum-xk=0}
\end{align}
which solves one of the lowest order Virasoro constraints on the string side, $C_{1}^{(0)}+C_{2}^{(0)}=0$ to be specific \cite{Chen:2005sb}.

\subsection{``Imaginary Root'' Case}

Given the double contour solution corresponding to folded string in
Lunin-Maldacena background, 
we can easily move on obtaining the solution for its circular
counterpart, this is the $\gamma$-deformed version of so-called
``imaginary root solution'' where all odd number of roots (before the
scaling limit) lie on the imaginary axis. Recall the analysis for this
class of solutions in the undeformed background
\cite{Beisert:2003xu,Beisert:2003ea,Arutyunov:2003rg}, 
here we do not have moding ambiguity, $n=0$ always; 
there are some odd number of Bethe roots congregate around the origin forming a constant density region (in the scaling limit they form a so-called ``condensate''), the remaining roots spread out on the either end of the constant density region. The Bethe string we have are two non-constant regions of equal length symmetrical under reflection about the real axis, joined by a constant region going through the origin. It is important to realize that despite the fact that all the roots lie on one continuous distribution, the presence of the constant condensate allows us to turn the situation into a two-cut problem again.
 Indeed as indicated in \cite{Kazakov:2004qf}, what one needs to do to relate the two-cut solutions for folded and circular is simply exchanging the role of the mode number $n$ and the condensate density $m$, or exchanging the $\A$- and $\B$-cycles.

Repeating the same analysis for the twisted spin chain as in \cite{Chen:2005sb} for the circular case, what is different from the undeformed case is that the condensate no longer congregate around origin but $2\pi\ga J$, the symmetry about the imaginary axis is again broken. However we still have two regions of non-constant Bethe roots density joined by a region of constant condensate. 
As it turns out in this case, (more in section
\ref{sec:outlines-Bethe}), 
we only have to exchange $x_{2}$ and $x_{3}$ to obtain the solution for this system, so the 
$\A$- and $\B$-cycle conditions (\ref{periods}) become, respectively,
\begin{align}\label{cycle-cond-IR}
2\pi m&=\oint_{\A}dp=2i\int_{x_{3}}^{x_{1}}\f{F(x) dx}{\sqrt{\ko{x_{1}-x}\ko{x-x_{3}}\ko{x-x_{2}}\ko{x-x_{4}}}}\,,\\
0&=\oint_{\B}dp=2\int_{x_{2}}^{x_{3}}\f{F(x) dx}{\sqrt{\ko{x_{1}-x}\ko{x_{3}-x}\ko{x-x_{2}}\ko{x-x_{4}}}}\,,
\end{align}
and the filling-fraction and the one-loop anomalous dimension are given by (here `IR' stands for `imaginary root')
\begin{align}
\al_{\rm IR}&=\f{1}{2}-\f{x_{1}x_{4}+x_{2}x_{3}}{4\sqrt{x_{1}x_{2}x_{3}x_{4}}}
-\f{\ko{x_{1}-x_{3}}\ko{x_{2}-x_{4}}}{4\sqrt{x_{1}x_{2}x_{3}x_{4}}}
\f{\E{\f{\ko{x_{2}-x_{3}}\ko{x_{1}-x_{4}}}{\ko{x_{1}-x_{3}}\ko{x_{2}-x_{4}}}}}{\K{\f{\ko{x_{2}-x_{3}}\ko{x_{1}-x_{4}}}{\ko{x_{1}-x_{3}}\ko{x_{2}-x_{4}}}}}\nonumber\\
&=\f{1}{2}-\f{x_{1}x_{4}+x_{2}x_{3}}{4\sqrt{x_{1}x_{2}x_{3}x_{4}}}
-\f{\ko{x_{1}-x_{2}}\ko{x_{3}-x_{4}}}{4\sqrt{x_{1}x_{2}x_{3}x_{4}}}
\f{\E{\f{\ko{x_{2}-x_{3}}\ko{x_{4}-x_{1}}}{\ko{x_{1}-x_{2}}\ko{x_{3}-x_{4}}}}}{\K{\f{\ko{x_{2}-x_{3}}\ko{x_{4}-x_{1}}}{\ko{x_{1}-x_{2}}\ko{x_{3}-x_{4}}}}}\,,
\label{fraction-IR}\\
\gamma_{\rm IR}&=\f{m^{2}}{128 \pi^{2}}\ko{\f{1}{x_{4}}-\f{1}{x_{2}}}\ko{\f{1}{x_{1}}-\f{1}{x_{3}}}\,
\f{\E{\f{\ko{x_{1}-x_{4}}\ko{x_{2}-x_{3}}}{\ko{x_{1}-x_{3}}\ko{x_{2}-x_{4}}}}}{\K{\f{\ko{x_{1}-x_{4}}\ko{x_{2}-x_{3}}}{\ko{x_{1}-x_{3}}\ko{x_{2}-x_{4}}}}}
-\f{m^{2}}{512\pi^{2}}\kko{\ko{\f{1}{x_{1}}+\f{1}{x_{4}}}-\ko{\f{1}{x_{2}}+\f{1}{x_{3}}}}^{2}\,,
\label{AD-IR}
\end{align}
where the integer $m$ represents the density of condensate in the deformed case.
From (\ref{ROS-c-LL}) and (\ref{fraction-IR}), we can see that the identifications (\ref{dictionary}) still hold for the circular case.  
Let us rewrite (\ref{AD-IR}) and (\ref{ENE-c-LL}) as
\begin{equation}
\gamma_{\rm IR}=\f{m^{2}}{2\pi^{2}}\,\E{1/q}\K{1/q}-\f{m^{2}}{512\pi^{2}}\kko{\ko{\f{1}{x_{1}}+\f{1}{x_{4}}}-\ko{\f{1}{x_{2}}+\f{1}{x_{3}}}}^{2}\,.
\end{equation}
and
\begin{equation}
E_{\rm circ}=\f{M^{2}}{2\pi^{2}}\,\K{1/q}\E{1/q}
+\f{M^{2}}{128\pi^{2}}\ko{\f{1}{x_{1}}+\f{1}{x_{4}}}\ko{\f{1}{x_{2}}+\f{1}{x_{3}}}\,,
\end{equation}
once again, upon imposing the Virasoro constraint (\ref{sum-xk=0}) and this time with $m=M$, we see they indeed match up.

\subsection{The Outlines of Bethe Strings\label{sec:outlines-Bethe}}

Analyzing the actual distribution of Bethe roots in the complex plane generally requires a numerical computation.
Nevertheless, once we take the scaling limit and reduce the problem of solving Bethe equations to a certain Riemann-Hilbert type problem, in principle, we can calculate the locations of the branch-points of the cuts from the given filling-fractions for each cut and the periods of cycles. However, in practice this is usually a very involved problem.
Here we instead discuss some plausible ans\"atz on how the roots would 
distribute and form Bethe strings in the scaling limit, 
our aim here is to illustrate these ans\"atze and complement our analysis in the previous subsections.

\begin{figure}[htb]
\begin{center}
\vspace{.7cm}
\hspace{-.0cm}\includegraphics[scale=0.64]{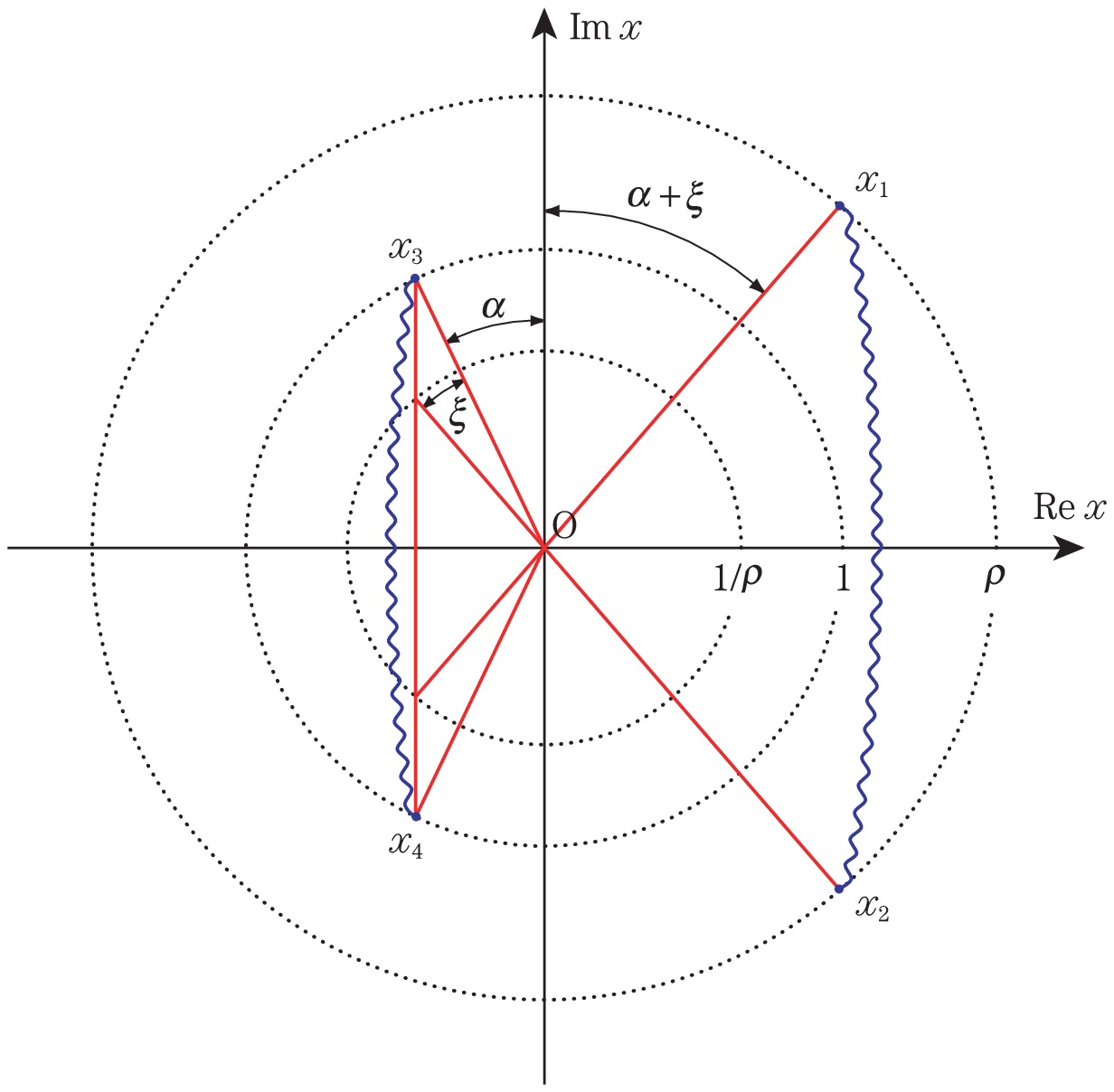}
\hspace{0.2cm}
\hspace{-.0cm}\includegraphics[scale=0.64]{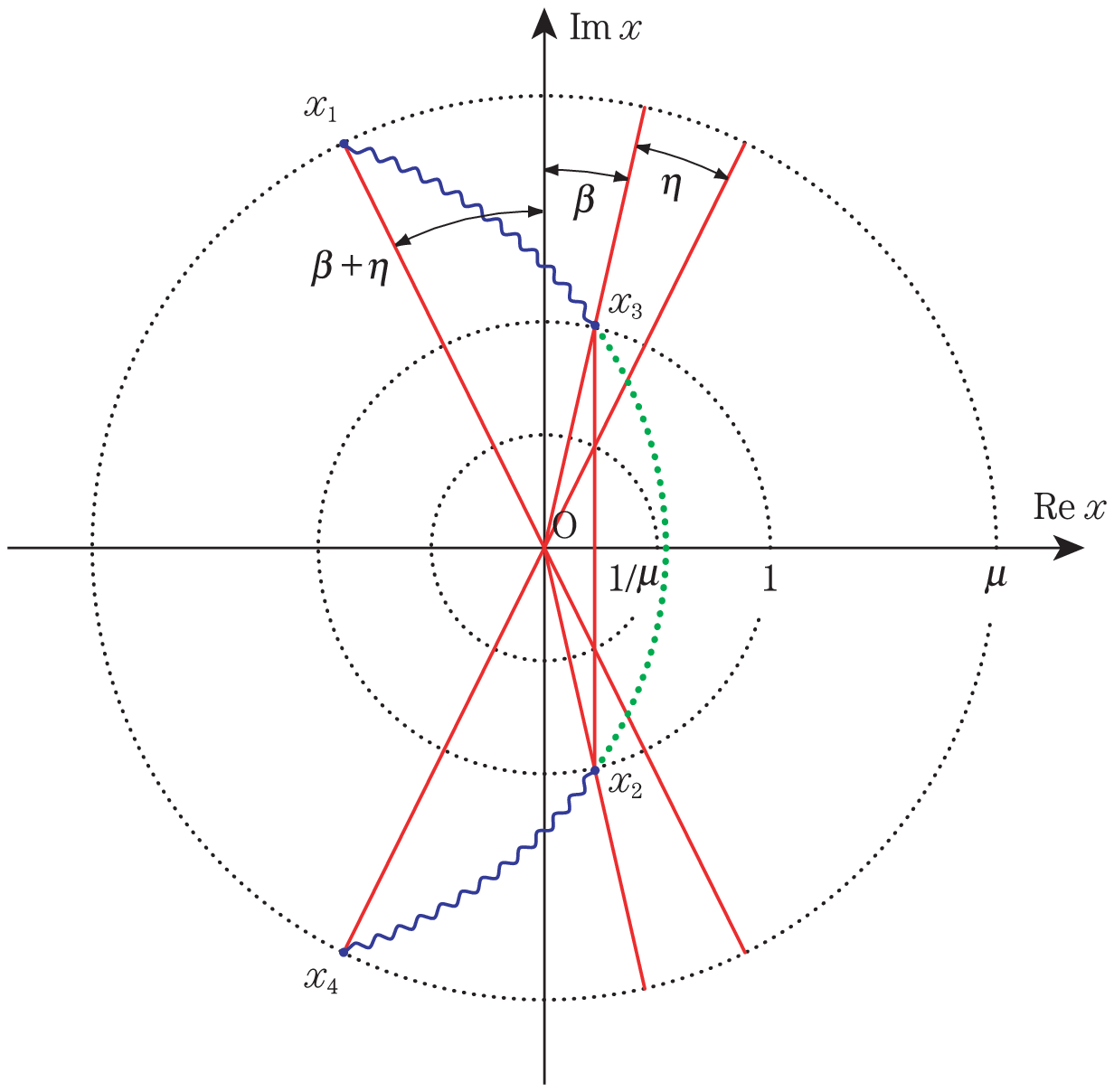}
\caption{\small{The ans\"atze for the outlines of Bethe string in the scaling limit.
The left figure is for the $\ga$-deformed version of a ``double contour'' solution and the right for the deformed ``imaginary root'' solution.
In the latter case, the dotted line joining $x_{2}$ and $x_{3}$ represents the ``condensate''.}}
\label{fig:beta-Bethe}
\end{center}
\end{figure}

Consider the following ans\"atz on the locations of the endpoints for the $\ga$-deformed version of ``double contour'' distribution, which we claim to be dual to the ``folded'' string in the deformed background with NR integrability:
\begin{equation}\label{ansatz-DC}
x_{1}=i\rho\, e^{-i\ko{\al+\xi}}\,,\quad 
x_{2}=x_{1}^{*}\,,\quad 
x_{3}=-\f{x_{2}}{\rho}\, e^{-i\xi}\,,\quad 
x_{4}=x_{3}^{*}\,,
\end{equation}
where $\rho$ is a real number greater than one, and the star $({}^{*})$ represents the complex conjugation.
The plot of this ans\"atz is in the left diagram of Fig.\ref{fig:beta-Bethe}.
The angles $\al$ (not to be confused with the filling-fraction) and $\xi$ are so drawn that they obey the relation $\rho \sin\al=\sin\ko{\al+\xi}$.
We can easily check the Virasoro condition (\ref{sum-xk=0}) is satisfied within the ans\"atz (\ref{ansatz-DC}).
This ans\"atz is also compatible with the so-called ``reality constraint'' which says the locations of the endpoints of the Bethe strings must be symmetrical about the real axis \cite{Kazakov:2004qf}.

In the limit of vanishing $\ga$ where $\xi\to 0$, we see the constraint $\rho \sin\al=\sin\ko{\al+\xi}$ requires $\rho\to 1$, and it is assumed that $\al$ goes to some finite value $\al_{0}$ to recover the symmetric two-cut distribution of the undeformed case.
If we denote the two cuts in the undeformed case as $\pm [a,b]$ and assume $x_{1}\to a$ and $x_{2}\to b$, it means $a/b=e^{2i\al_{0}}$ is satisfied.

Even for the $\gamma$-deformed case, we can apply the trick of solving this system for the negative filling ratio as in \cite{Beisert:2003xu}, this amounts to rotating the two cuts to the real axis. This useful trick allows us in the scaling limit, to neglect the actual shapes of the cuts but only care about the end-points, we end up with the two asymmetrical cuts on real axis used in \cite{Chen:2005sb}.

Using the identification (\ref{dictionary}), we can rewrite the moduli parameters $y_{0}$ and $y_{\pm}$ on the string side in terms of the end points on the spin chain side and obtain
\begin{equation}\label{translation-DC}
y_{+}=\f{1}{2}+\f{1}{4}\ko{\rho+\f{1}{\rho}}\,,\quad
y_{-}=\f{1}{2}-\f{1}{2}\cos\xi\,,\quad 
y_{0}=\f{1}{2}+\f{1}{2}\cos\ko{2\al+\xi}\,.
\end{equation}
We can see they indeed satisfy $0<y_{-}<y_{0}<1<y_{+}$, corresponding to (c) of Fig.\,\ref{fig:transition}.
Comparing these with the definitions of $\theta_{-}$ and $\theta_{0}$ that determines the range of oscillation for the folded string, we can deduce $\theta_{-}\eq \xi$ and $\theta_{0}\eq \pi-(2\al+\xi)$.
We also see the short-string limit where both $\theta_{-}$ and $\theta_{0}$ tend to zero indeed corresponds to the short Bethe-string limit of $\al\to \pi/2$ and $\xi\to 0$, showing this ans\"atz has a smooth BMN limit.

Next let us turn to the case of the $\ga$-deformed version of ``imaginary root'' solution.
Again imposing the reality constraint, we propose the following ans\"atz:
\begin{equation}\label{ansatz-IR}
x_{1}=i\mu\, e^{i\ko{\be+\eta}}\,,\quad 
x_{2}=-\f{x_{1}}{\mu}\, e^{-i\eta}\,,\quad 
x_{3}=x_{2}^{*}\,,\quad
x_{4}=x_{1}^{*}\,.
\end{equation}
This is illustrated in the right diagram of Fig.\,\ref{fig:beta-Bethe} with condensates located between $x_{2}$ and $x_{3}$.
Here $\mu$ is a real number greater than one, and $\beta$ and $\eta$ satisfy the relation $\mu\sin\beta=\sin\ko{\beta+\eta}$ as can be seen from Fig.\ref{fig:beta-Bethe}, so that the Virasoro condition (\ref{sum-xk=0}) is satisfied.
In the limit of vanishing $\ga$, both $\eta$ and $\be$ tend to zero simultaneously, while $\mu$ goes to some finite value $\mu_{0}$, which can be written as $\mu_{0}\eq t/s$ with the two cuts located in the range $\pm [is, it]$. 
To obtain the filling-fraction and the spin-chain energy, it is again convenient to apply the same trick as in the double contour solution, and rotate the entire continuous curve to the real axis, similar calculations then give (\ref{fraction-IR}) and (\ref{AD-IR}). 
The moduli for the circular string in the deformed background can be written as
\begin{equation}\label{translation-IR}
y_{+}=\f{1}{2}+\f{1}{2}\cos\ko{2\be+\eta}\,,\quad
y_{-}=\f{1}{2}-\f{1}{2}\cos\eta\,,\quad 
y_{0}=\f{1}{2}+\f{1}{4}\ko{\mu+\f{1}{\mu}}\,,
\end{equation}
which satisfies $0<y_{-}<y_{+}<1<y_{0}$, corresponding to (f) of Fig.\,\ref{fig:transition} and reduce to (c) of Fig.\,\ref{fig:transition} in the vanishing-$\gamma$ limit.

Finally let us consider the rational limit.
In this limit, the filling-fraction tends to $\f{1}{2}$ (i.e., ``half-filling''), and the outer two of the four branch-points, $x_{1}$ and $x_{4}$, go to infinity and reduce the two-cut problem to the one-cut, with inner two, $x_{2}$ and $x_{3}$ remaining at finite distance from the origin.
In view of (\ref{translation-IR}), the rational limit $\mu\to \infty$ means sending $y_{0}$ to $\infty$ on the string side, and thus we reproduce the observation made earlier.
This limit also imply the vanishing of angle $\be$, which we can see from the constraint $\mu\sin\beta=\sin\ko{\beta+\eta}$, hence the inner two branch-points in fact approach the imaginary axis, namely at $\pm i/(m\pi)$.
One can see there is no BMN limit in this circular case.

\section{Summary and Outlook\label{sec:conclusion-discussions}}

In this paper we presented further corroborative evidences for the spin-chain/spinning-string duality in the Lunin-Maldacena background whose gauge theory dual is the Leigh-Strassler deformation of ${\mathcal{N}}=4$ SYM.
We restricted ourselves to the case of real deformation parameter $\gamma$ on both sides of the correspondence.
On the string sigma model side, we considered the large-spin limit of the $SU(2)_{\ga}$ sector and demonstrated how the Landau-Lifshitz action can be obtained for this sector.
The Landau-Lifshitz action was shown to agree with the coherent state path integral of the twisted Heisenberg spin-chain Hamiltonian associated with the $SU(2)_{\ga}$ sector of the gauge side, reproducing the observation made in \cite{Frolov:2005ty}.

We then made it clear that, in the Lunin-Maldacena background, spinning string solutions cannot arise from the well-known Neumann system, instead they should be the solutions to the more general NR system. Other than usual rotating motions, the strings generally also extend in the directions associated with the two of the $U(1)\times U(1)\times U(1)$ Cartan subgroup of the $SU(4)$ R-symmetry group for ${\mathcal{N}}=4$ SYM, which is preserved under the deformation.
Analogous to the case of the Neumann integrable system in the undeformed background, we showed that there are two kinds of solitonic solutions allowed in the Lunin-Maldacena background, and the energy-spin relation in the large spin limit were calculated for both cases.
These solutions typically had the profiles like Fig.\ref{fig:string-on-LM-nutation}, and for specific choices of the winding and oscillating numbers, they were shown to reduce to an eccentric version of folded or circular solutions of the usual Neumann system. The folded solution of this special case turned out to be the one obtained in \cite{Chen:2005sb}.

From the spin chain side, the $\ga$-deformed version of the ``imaginary root'' solution was analyzed as its counterpart of the deformed ``double contour'' solution studied in \cite{Chen:2005sb}.
The filling-fractions and the anomalous dimensions for both cases were calculated and were shown to be related via certain analytical continuation.
Furthermore, we proposed explicit ans\"atze for the distributions of Bethe roots, both for the deformed versions of double contour and imaginary root solutions. The endpoints of cuts were shown to be compatible with the necessary conditions such as reality condition, and to have the expected behaviors in various limits.
Remarkably, the moduli parametrizing the spinning string solutions in Lunin-Maldacena background were identified with the geometrical quantities in the associated Bethe string ans\"atz.
These identifications allowed us to explicitly demonstrate the exact matching between the string energies and the one-loop anomalous dimensions calculated from the twisted spin chain.

Here we outline few possible interesting extensions. 

It would be interesting to investigate the matching of higher conserved charge from both sides of the correspondence in the deformed set-up, following the work of \cite{Arutyunov:2003rg, Arutyunov:2004xy,Arutyunov:2005nk}, and we hope to report more on this in the near future \cite{CO2:2006xx}.

It would be also interesting to extend our results to the case of larger symmetry groups.
The duality between the $SU(3)_{\ga}$ sectors of both string and gauge theory was examined and a quite non-trivial agreement at the level of effective actions was achieved \cite{Frolov:2005iq}.
With the possible help of the explicit solutions to these two approaches, i.e.  the Bethe ans\"atz technique and the Landau-Lifshitz approach, we will be able to perform more explicit non-trivial tests to their correspondence in the deformed background.

Extension to the higher-loop cases, that is higher order in the effective coupling $\f{\lambda}{J^{2}}$, can also be meaningful.
The Landau-Lifshitz reduced action for the complex $\be$-deformed $SU(2)$ sector at the two-loop level was explicitly presented in \cite{Frolov:2005ty}, so in principle the problem of computing conserved charges would be accessible.
Moreover, it would also be interesting, albeit potentially difficult, to come up with a deformed version of two-loop spin chain and modified ans\"atz as in \cite{Serban:2004jf}. The combination of these results should lead to the highly non-trivial explicit matchings of the higher conserved charges at two-loop order (c.f.\cite{Arutyunov:2004xy}), we shall also explore this in future.

Finally, it is also important to study the finite-size (subleading in $1/J$-expansion) corrections to the energy/anomalous-dimension in the context of deformed theories.
This line of study has been partly carried out in \cite{Frolov:2005ty} for the case of rational circular solutions and the agreement including the subleading order has been shown.
For the elliptic solutions, however, the problem of calculating the $1/J$-corrections does not seem easy even in the undeformed case, and the problem of checking the duality including this finite-size corrections remains a challenging task.

\subsection*{Acknowledgments}

We would like to thank N.~Dorey, Y.~Imamura and Y.~Nakayama for reading and commenting the draft, and  K.~Ideguchi, S.~P.~Kumar and Y.~Matsuo for their valuable comments and discussions. HYC would like to thank the financial support of St.~John's college, Cambridge through a Benefactors Scholarship.

\appendix
\section*{Appendix}

\section{Complete Elliptic Integrals\label{app:elliptic}}

Our convention for the complete elliptic integrals of the first, second and third kind are as follows:
\begin{alignat}{3}
\K{r} &\equiv \int_{0}^{1} \frac{d x}{\sqrt{\ko{1-x^2}\ko{1-rx^2}}}&{}&=\int_{0}^{\pi/2} \frac{d \varphi}{\sqrt{1-r\sin^{2}\varphi}}\,,\\
\E{r} &\equiv \int_{0}^{1} d x\,\sqrt{\frac{1-rx^2}{1-x^2}}&{}&=\int_{0}^{\pi/2} d \varphi\,\sqrt{1-r\sin^{2}\varphi}\,,\\
\PP{q,r} &\equiv \int_{0}^{1} \frac{d x}{\ko{1-qx^{2}}\sqrt{\ko{1-x^2}\ko{1-rx^2}}}&{}&=\int_{0}^{\pi/2} \frac{d \varphi}{\ko{1-q\sin^{2}\varphi}\sqrt{1-r\sin^{2}\varphi}}\,.
\end{alignat}
The integral formulae for $a>b>c$ listed below are useful in the calculation in the main text:
\begin{align}
&\int_{c}^{b}\f{dx}{\sqrt{\ko{a-x}\ko{b-x}\ko{x-c}}}=\f{2}{\sqrt{a-c}}\, \K{\komoji{\f{b-c}{a-c}}}\,,\label{ellip-int-formula-1}\\
&\int_{c}^{b}\f{x dx}{\sqrt{\ko{a-x}\ko{b-x}\ko{x-c}}}=\f{2a}{\sqrt{a-c}}\, \K{\komoji{\f{b-c}{a-c}}}-2\sqrt{a-c}\, \E{\komoji{\f{b-c}{a-c}}}\,,\label{ellip-int-formula-2}\\
&\int_{c}^{b}\f{dx}{x\sqrt{\ko{a-x}\ko{b-x}\ko{x-c}}}=\f{2}{c\sqrt{a-c}}\, \PP{\komoji{-\f{b-c}{c},\f{b-c}{a-c}}}\,.\label{ellip-int-formula-3}
\end{align}

\section{The Riemann-Hilbert Problem for Twisted Spin-Chain\label{sec:RH-beta}}

In this appendix, we will make a brief review on some relevant aspects of the Riemann-Hilbert problem associated with the field theory.
In \cite{Frolov:2005ty}, the analysis of the complex $\be$-deformed $SU(2)$ sector was performed for a generic case at the two-loop level in $\lambda$.
Below we will summarize the results of \cite{Frolov:2005ty}, restricting ourselves to the real part of $\be$ denoted as $\gamma$.
For details, see the earlier works \cite{Frolov:2005ty,Chen:2005sb} and the references therein.

For the $SU(2)_{\ga}$ sector of gauge theory in the scaling limit, the problem of solving a set of Bethe equations becomes a particular Riemann-Hilbert problem with two cuts on the associated Riemann surface.
The anomalous dimension (or the energy of the spin-chain system) can be described by the periods of two cycles and some suitably defined elliptic moduli.
As we mentioned in the main text, we assume $J\eq J_{1}+J_{2}$ and $J_{1,2}$ large, while $\ga J$ stays finite.

We shall concentrate on the situation where the two cuts $\C_{\pm}$ have shifted mode numbers $n_{\pm}\eq \pm n+\ga J$ respectively, where $n$ is an integer, and there is a ``condensate'' of density $m=\widetilde m+\gamma J$ between them. Here $\widetilde m$ can be regarded as the non-integer valued condensate density in the undeformed background, so that combined with $\gamma J$, we obtain the physical integer valued condensate density $m$ in the deformed background.
The integers $n$ and $m$ are related to the so-called $\A$- and $\B$-cycles of the two-cut problem, and one of the most important features in this $\ga$-deformed theory is that the effect of the deformation only results in the shifts of the periods of each cycle \cite{Frolov:2005ty}.

Let us denote the number of Bethe roots as $M\eq J_{2}$, and introduce the filling fraction $\alpha=M/J$.
As usual, define the resolvent as 
\begin{equation}\label{def:G}
G(x)\eq \f{1}{J}\sum_{j=1}^{M}\f{1}{x-x_{j}}\,.
\end{equation}
Furthermore, we can introduce the quasi-momentum $p(x)$, which is related to the resolvent as
\begin{equation}\label{def:p}
p(x)=G(x)-\f{1}{2x}\,,
\end{equation}
so that the ``twisted'' Bethe equation can be rewritten in the form,
\begin{equation}\label{Bethe-p}
p(x+i\epsilon)+p(x-i\epsilon)=2\pi n_{\pm}\,,\qquad x\in\C_{\pm}\,.
\end{equation}
The momentum condition can also be written as
\begin{equation}\label{momentum-cond}
-\oint_{\C_{+}\,\cup\,\C_{-}}\f{dx}{2\pi i}\f{G(x)}{x}=2\pi m
\end{equation}
with the condensate density.
We can see the equations (\ref{Bethe-p}) and (\ref{momentum-cond}) have exactly the same structures as in the original {\Nf} case, and the only difference is the shifts of numbers such that $\pm n\mapsto n_{\pm}=\pm n+\ga J$ and $\widetilde m\mapsto m=\widetilde m+\ga M$.

Following \cite{Kazakov:2004qf,Chen:2005sb,Okamura:2005cj}, we now introduce the genus one Riemann surface by the following elliptic curve,
\begin{equation}
y^{2}=\prod_{k=1}^{4}\ko{x-x_{k}}=x^{4}+c_{1}x^{3}+c_{2}x^{2}+c_{3}x+c_{4}\,,
\end{equation}
which is endowed with a meromorphic differential $dp(x)$ of the form
\begin{equation}
dp(x)=\f{dx}{\sqrt{\prod_{k=1}^{4}\ko{x-x_{k}}}}\sum_{k=-1,0,1}a_{k}x^{k-1}\,.
\end{equation}
By demanding $dp(x)=\f{dx}{2x^{2}}+\ord{1}$ as $x\to 0$, we can easily determine the coefficients $a_{-1}$ and $a_{0}$ as 
\begin{equation}
a_{-1}=\f{1}{2}\sqrt{c_{4}}=\f{1}{2}\sqrt{\prod_{k=1}^{4}x_{k}}\,,\qquad 
a_{0}=\f{c_{3}}{4\sqrt{c_{4}}}=-\f{1}{4}\sqrt{\prod_{k=1}^{4}x_{k}}\ko{\sum_{k=1}^{4}\f{1}{x_{k}}}\,.
\end{equation}
whereas $a_{1}$ can be fixed by the normalization condition as
\begin{equation}
a_{1}=\f{1}{2}-\al\,.
\end{equation}
The periods for the so-called $\A$- and $\B$-cycles are given by integer valued $n$ and $m$, 
\begin{equation}\label{periods}
\oint_{\A}dp=2\pi m\qquad \mbox{and}\qquad \oint_{\B}dp=2\pi \ko{n_{+}-n_{-}}=4\pi n\,.
\end{equation}
Note that the $\B$-period is not affected by the deformation, since it is given by the difference of two numbers $n_{\pm}$, which are equally shifted by the deformation \cite{Frolov:2005ty}.
For the deformed double contour case, we set $m=0$ whereas for the deformed imaginary root case $n=0$, and the $(n,m)=(1,0)$ case was studied in \cite{Chen:2005sb}.

Finally let us summarize the counting of free parameters and the constraints in the system.
At first there seems to be four complex parameters $x_{1},\dots, x_{4}$ and one real parameter $a_{1}$ to be fixed.
But in light of the so-called ``reality constraint'' \cite{Kazakov:2004qf} as well as the Bethe equation $p(\infty_{+})-p(\infty_{-})=2\pi n_{+}$ (where $\infty_{\pm}$ are two points at infinity for each sheet of the Riemann surface), we are left with four real degrees of freedom.
These can be fixed by the $\A$- and the $\B$-cycle conditions (\ref{periods}), and giving the filling fractions $\al_{\pm}$ for each cut $\C_{\pm}$, which read $\al_{+}=\al_{-}=\al/2$ in our case.



\end{document}